\documentclass[letterpaper]{aa}  
\usepackage[dvips]{graphicx}
\usepackage{txfonts}
\usepackage{psfig,times} 
\def\ergs{erg\,cm$^{-2}$\,s$^{-1}$}
\def\erglu{erg\,s$^{-1}$}
\def\s5{S5~0716+71}
\begin{document}
   \title{Disentangling the 
synchrotron and Inverse Compton variability in the
   X-ray emission of the intermediate \\
   BL Lac object \s5}

%   \subtitle{} 

   \author{E. Ferrero \inst{1},
           S. J. Wagner\inst{1},
           D. Emmanoulopoulos\inst{1},
           L. Ostorero\inst{1}
          }

   \offprints{E. Ferrero; e-mail: eferrero@lsw.uni-heidelberg.de}

   \institute{\inst{1}Landessternwarte Heidelberg, K\"onigstuhl - 69117,
 Heidelberg,
   Germany  
             }

   \date{Received ?; accepted ?}

%\abstract{}{}{}{}{} 
% 5 {} token are mandatory
 
  \abstract
   {The possibility to detect simultaneously in the X-ray band 
   the synchrotron and Inverse Compton (IC) 
   emission of intermediate BL Lac
   objects offers the unique opportunity to study contemporaneously
   the low- and high-energy tails of the electron distribution in the jets of 
   these sources.}
   {We attempted to disentangle the X-ray spectral variability 
   properties 
   of both the low- and high-energy ends of the synchrotron
   and  Inverse Compton emission of the intermediate BL Lac object \s5. }
   {We carried out 
    spectral, temporal and cross-correlation analyses of the data from a long
   XMM-{\it Newton} pointing of \s5 and we compared
   our findings with previous results from past X-ray observations. }
   {Strong variability was detected during the XMM exposure.
    Both the synchrotron and Inverse Compton components were found to vary 
    on time scales of hours, implying a size of the emitting region
of $R\la 0.7\delta /(1+z)$ light-hours.  
The synchrotron emission
   was discovered to become dominant during
   episodes of flaring activity, 
   following a harder-when-brighter trend. Tight
   correlations were observed between variations in different energy bands.
   Upper limits on time lags 
   between the soft and hard X-ray light curves are of the order of a few
   hundred seconds.}
   {}


   \authorrunning{E. Ferrero et al.}
   \titlerunning{XMM-Newton observation of \s5}
   \maketitle
%
%________________________________________________________________

\section{Introduction}
\label{intro}

BL Lac objects, as well as Flat Spectrum Radio Quasars (FSRQ),
 are the most extreme 
radio-loud Active Galactic Nuclei (AGN). One of their most striking
properties is the strong and rapid
variability at all wavelengths.  The observed radiation from BL Lac objects
is commonly believed to be dominated by non-thermal emission 
from a jet pointing roughly towards the observer and 
moving at relativistic velocity 
(Blandford \& Rees 1978). The Spectral Energy
Distributions (SED) of BL Lacs are typically 
characterized, in $\nu $-$\nu F_{\nu }$ 
representation,  by two broad bumps. The low-energy one, peaking 
at frequencies from the IR up to the soft X-ray band, is 
usually interpreted as 
synchrotron emission by a population
of relativistic electrons in 
the jet. The high-energy one, peaking at gamma-ray energies, 
 is supposed to be Inverse Compton (IC) radiation
by the same population of electrons, scattering either the synchrotron
photons themselves (Synchrotron Self-Compton, SSC; e.g. Jones, O'Dell \& Stein
 1974; Maraschi et al. 1992;
Kirk \& Mastichiadis 1997) or external photons from the surrounding environment
(External Compton,
EC; e.g. Dermer et al. 1992; Sikora et al. 1994). \\
BL Lac objects were originally divided into radio-selected (RBL) and 
X-ray-selected (XBL), according to the waveband of their discovery
      (Ledden \& O'Dell 1985). 
A more physical classification (Padovani \& Giommi 1995)
separates BL Lacs 
depending on whether the synchrotron peak is situated at IR-optical frequencies
(Low-energy-peaked BL Lacs, LBL) or in the UV/soft X-ray band 
(High-energy-peaked BL
Lacs, HBL). \\
HBL are the brightest BL Lacs in the X-ray band and thus they are
the best 
studied ones at these energies. 
The soft ($\Gamma \ga 2$) and
continuously steepening towards higher energies 
X-ray spectra of HBL (Perlman et al. 2005) are
commonly interpreted in terms  of 
synchrotron radiation from the high-energy
tail of the electron distribution. This tail is expected to be very sensitive 
to particle
acceleration and cooling time scales and thus to be mostly affected by rapid 
and
strong variability (Kirk \& Mastichiadis 1997). 
X-ray observations of the closest and X-ray brightest 
HBL (PKS 2155-305: e.g. Chiappetti et 
al. 1999; Edelson et al. 2001; Zhang et al. 2005; 
MRK 421: e.g. Takahashi
et 
al. 1996; Fossati et al. 2000; Brinkmann et al. 2005; MRK 501: e.g. Pian et al. 1998; Tavecchio et al. 2001), have indeed
revealed strong variability, characterized by large amplitude
 variations, both on
long and short time scales  
(fastest flux changes of  $\sim 2$ on time scales 
of the order of a few hours; Sembay et al. 1993; Zhang et al. 2002). The
variability amplitude has been found to be correlated with 
energy (Sembay et al. 2002; Zhang et al. 2005; Gliozzi et al. 2006), 
in agreement with the hypothesis that the hardest synchrotron 
radiation is produced by the most energetic electrons with the smallest 
cooling time scales.
Flux increases are typically accompanied by spectral
hardening, with the remarkable
exception of the July 1997 flare of MRK 501, which
 exhibited an opposite behaviour (Lamer \& Wagner 1998). 
However, from a recent reanalysis of the 1997 RXTE data, Gliozzi et al. (2006)
concluded that this was a spurious result.  
Shifts  of the synchrotron peaks, 
up to two orders
of magnitude in the case of MRK 501, towards higher energies
have been observed during extreme flaring activity (Pian et al. 1998; Giommi
 et al. 2000; Tavecchio et al. 2001). Light curves in different X-ray bands have been
found to be well correlated, often with delays of the order of a few hours.
However, independent variability in different X-ray bands has also been
 observed during the RXTE observations of MRK 501 in July 1997 (Lamer \&
 Wagner 1998).
The length of the lags, when detected, appears to change from flare to
flare. 
Variations at softer energies usually lag behind those at harder energies
(soft lags), although 
the opposite behavior (hard lags) 
 has also been claimed 
(PKS 2155-305: Zhang et al. 2006; MRK 421: Fossati et al. 2000).  
Time lags lead to spectral changes with flux characterized, in  
spectral index vs. intensity ($\alpha -I$) plots,  by
clockwise (soft lags) or counter-clockwise (hard lags) loop patterns, 
effectively observed in some cases (e.g. Takahashi et al. 1996; Zhang et
 al. 2002; Ravasio et al. 2004; Brinkmann et al. 2005).
The sign of the time lags and the paths traced in the
 $\alpha -I$ plane have been directly related to 
differences
in the acceleration and cooling time scales of the system for the energy bands
considered 
(Kirk, Rieger \& Mastichiadis 1998). This in turn has been used to impose 
constraints
 on some of the physical
parameters of the emitting regions, such as the magnetic field and the Lorentz
factors  of the particles. \\
On the other hand, the X-ray emission of LBL is believed to originate mainly
from IC scattering of seed photons
by the low-energy tail of the electron population, although the synchrotron
emission of the high-energy particles might still contribute significantly. 
For a few objects  with synchrotron
peaks located around $10^{14}-10^{15}$ Hz, known as intermediate BL Lac
objects (IBL), the X-ray observations have clearly
detected the turning point 
of their SED, where the synchrotron and IC components intersect
(ON 231: Tagliaferri et al. 2000; BL Lacertae: Tanihata
et al. 2000; Ravasio et al. 2002; \s5: Cappi et al. 1994; Wagner et al. 1996;
Giommi et al. 1999; 
Tagliaferri et
al. 2003; AO 0235+16: Raiteri et al. 2006; OQ 530: Tagliaferri et al. 2003). 
X-ray studies of both flux and spectral variability of LBL/IBL are thus
able, in principle, to convey information on the physical conditions of 
both the low- and high-energy
electrons simultaneously; in the case of HBL this goal can be achieved only
through contemporaneous multi-frequency observations. 
However, due to their lower X-ray luminosities as compared with HBL, 
only a few
X-ray studies of LBL/IBL have been carried out so far. 
In general, the X-ray variability of known LBL/IBL
shows similar characteristics as HBL, both in terms of 
amplitude and time scales. 
However, contrary to HBL, the variability
amplitude seems to decrease at harder energies (Giommi et al. 1999; Ravasio et
 al. 2002), where the IC
component becomes dominant. This finds a natural explanation in the longest 
cooling time scales of the lowest energy electrons responsible for the IC
emission. 
Short time scale ( $\la $ hours) variability seems to be present only 
in the synchrotron component, whereas the IC
emission appears to vary on longer time scales ($\sim $ days). 
 To our knowledge, no time lags between different X-ray bands 
or loops in the  $\alpha -I$ plane
have been
reported so far for LBL/IBL, although they might be expected (B\"ottcher \& Chiang
2002). \\
In this paper we present
the analysis of an archival XMM-{\it Newton} observation of \s5, lasting $\sim 59$ ks. 
It is the longest uninterrupted, highest  signal-to-noise ratio
X-ray observation
performed so far for this source, 
 thus providing the possibility to
disentangle with great accuracy 
the relative contributions, within the XMM band, of the synchrotron and IC 
components, 
 and to determine the variability properties of both of them.\\
Recently, a separate study of the same XMM data
set was presented by 
Foschini et al. (2006) during the submission process
of our manuscript. 
Their results are briefly compared with ours
in Sect. \ref{discussion}. \\
 The outline of the paper is the following: in Sect. \ref{background} we give more details on the source and we summarize
briefly the results from past X-ray observations;  
in Sect. \ref{dataproc} we describe the data processing; in
Sects. \ref{spectral} and \ref{timing} 
  we give, respectively, the results of the spectral and timing analysis;
  in Sect. \ref{timespectral} we analyze the spectral variability 
of the source; in Sect. \ref{optical} we examine
  the Optical Monitor data; in Sect. \ref{discussion} we summarize and
discuss the results, and in Sect. \ref{conclusions} we give our conclusions.

%__________________________________________________________________

\section{Past X-ray observations}
\label{background}

\s5 is one of the brightest and most active BL Lac objects in the sky. 
It was initially discovered in the 5 GHz Bonn-NRAO radio survey and was
included in the S5 catalog of strong ($S>1$ Jy), flat ($\alpha \le 0.5$, 
$S\propto \nu ^{-\alpha}$) radio sources with declination $\delta \ge
70^{\circ }$ (K\"uhr et al. 1981). 
In spite of its optical brightness
($m_{\rm V} \sim 13.0$), its spectrum does not show any emission or absorption
feature which, together with its high linear polarization,
 led to its classification as a BL Lac object (Biermann et al. 1981).  An
 upper limit on the redshift of $z>0.3$ was estimated from the lack of any
 detection of the host galaxy in optical images (Wagner et
 al. 1996). Recently, Sbarufatti et al. (2005) proposed a new lower limit of
 $z>0.52$ using HST images. \s5 is a well known intra-day variable (IDV)
 source with a very high 
 variability duty cycle and it has been intensively
 studied at all frequencies since its discovery. In particular, \s5 has been
 the target of a number of multi-frequency campaigns (e.g. Wagner et al. 1990;
 Quirrenbach et al. 1991; Wagner et al. 1996; Ostorero et al. 2006).
The synchrotron peak of its SED falls around $10^{14}$ Hz 
(Wagner \& Witzel 1995; Ostorero et al. 2006).\\
\s5 has also been observed several times 
 by various X-ray telescopes and some of 
the results are summarized in Table
 \ref{pastx}.  Biermann et al. (1992) were the first to report an X-ray flux
 from a HEAO-A observation in 1977. The source was then observed by the 
 {\it Einstein} satellite in 1979 and 1980; however, the counts were  
too low to determine a spectrum and only a flux estimate could be given 
(Biermann et al. 1981). A spectral and temporal analysis was
possible for the first time with ROSAT, which observed \s5 with the PSPC in 
March 1991 (Wagner 1992; Cappi et
al. 1994; Urry et al. 1996; Wagner et al. 1996). These observations
suggested that two distinct spectral components are necessary to
account for the X-ray spectrum in the 0.1--2.4 keV band and these components
 were
 interpreted
as synchrotron and IC emission, respectively. The same data 
clearly revealed for the first time in this source
flux variability with associated  
spectral variations.  In 1994, \s5 was
observed by ASCA as part of a series of multi-wavelength campaigns on a 
                      sample of
blazars (Kubo et al. 1998). ASCA confirmed 
the existence of a spectral flattening with
increasing energies, in agreement with the ROSAT results.
 During a multi-frequency campaign in 1996, \s5 was the
target of simultaneous ROSAT-HRI (0.1--2.4 keV) and RXTE (2--10 keV)
pointings. No correlation could be established 
between the corresponding light curves, a result which, given the different
energy range of the two
instruments, pointed at the presence of
a soft and hard components with different variability properties
(Otterbein et al. 1998). 
 Finally, \s5 was
observed by {\it Beppo}SAX in 1996, 1998 (Giommi et al. 1999) and 2000
(Tagliaferri et al. 2000). In 1996 and 1998 the source was found in a faint
state and the spectral modeling required two components, similarly to 
previous X-ray
observations. Short-term variability of the soft component below $\sim 3$ keV
was detected; the hard component above $\sim 5$ keV appeared to be
 more stable,
although some long-term variability seemed to be present
from the comparison between the 1996 and 1998 data.
In 2000 the
 source was caught in its highest state and the soft slope was the largest of
 the three {\it Beppo}SAX observations, suggesting a softer-when-brighter
trend, opposite to that usually observed for BL Lacs. The 2000 data confirmed the essential absence of variability above $\sim
 3 $ keV. \\

\begin{table*}
\small
\tabcolsep1ex
\caption{\label{pastx} Results from past X-ray observations (see
  Sect. \ref{background} for further details). }
\begin{tabular}{ccccccccc}
\noalign{\smallskip} \hline \noalign{\smallskip}
\hline
\multicolumn{1}{c}{Date} & \multicolumn{1}{c}{Instrument} & 
\multicolumn{1}{c}{$\Gamma _{\rm soft}$} & \multicolumn{1}{c}{$E_{\rm break}$ (keV)} &
\multicolumn{1}{c}{$\Gamma _{\rm hard}$} & \multicolumn{1}{c}{Band (keV)} 
& \multicolumn{1}{c}{$F$ (\ergs)} & \multicolumn{1}{c}{$F_{\rm 1~keV}$ (\ergs
  Hz$^{\rm -1}$)} & \multicolumn{1}{c}{Ref.} \\
\noalign{\smallskip} \hline \noalign{\smallskip}
2 Oct 1977 & HEAO-A & (1.5)$^{*}$ & & & 0.25--25 & & $1.28\times 10^{-29}$ & (2)\\
29 Aug 1979 & {\it Einstein } & (1.5)$^{*}$ & & & 0.2--3.5 & $2.89\times 10^{-12}$
& $0.41\times 10^{-29}$ & (1) \\ 
19 Oct 1979, 9 Mar 1980  & {\it Einstein } & (1.5)$^{*}$ & & & 0.2--3.5 & 
& $0.31\times 10^{-29}$ & (2) \\
8/11 Mar 1991 (low state) & ROSAT-PSPC & $3.99^{\rm +4.21}_{\rm -3.81}$ & & 
$2.25^{\rm +2.65}_{\rm -1.25}$ & 0.1--2.4 & $3.73\times 10^{-11}$ & 
$1.03\times 10^{-29}$ & (3) \\
8/11 Mar 1991 (high state) & ROSAT-PSPC & (3.99)$^{*}$ & & 
(2.25)$^{*}$ & 0.1--2.4 & $8.45\times 10^{-11}$ & 
$2.98\times 10^{-29}$ & (3) \\
16/19/21 Mar 1994 & ASCA & & & & 2--10 & $1.30\times 10^{-12}$ & & (4)\\
24 Mar--22 Apr 1996 & ROSAT-HRI & & & & & & & (5) \\
6--22 Apr 1996 & RXTE & & & 3.1$\pm$0.1 & 2--10 &  &  & (5) \\
14 Nov 1996 & BeppoSAX & 2.7$\pm$0.3 & 2.3$\pm$0.4 & 1.96$\pm$0.15 & 2--10 &
$1.4\times 10^{\rm -12}$ & & (6) \\
7 Nov 1998 & BeppoSAX & 2.3$\pm$0.4 & 2.8$\pm$0.8 & 1.73$\pm$0.18 & 2--10 &
$2.6\times 10^{\rm -12}$ & & (6) \\
30/31 Oct 2000 & BeppoSAX & $3.40^{\rm +0.4}_{\rm -0.3}$ & & 
$1.60^{\rm +0.25}_{\rm -0.35}$ & 2--10 & $3.3\times 10^{\rm -12}$ & 
$1.80\times 10^{-29}$ & (7) \\

\noalign{\smallskip}\hline
\end{tabular}
\medskip

* Fixed photon index. 
(1) Biermann et al. (1981),
(2) Biermann et al. (1992),                                                  
(3) Cappi et al. (1994),
(4) Kubo et al. (1998),
(5) Otterbein et al. (1998),
(6) Giommi et al. (1999),
(7) Tagliaferri et al. (2003).

\end{table*}

%__________________________________________________________________
                                                                                                                                                                 
\section{The XMM-Newton observation}
\label{dataproc}

\s5 was observed by XMM-{\it Newton} (Jansen et al. 2001) 
for $\sim $ 59 ks on April 4--5, 2004 (Obs. ID 0150495601, Rev. 791,
PI G. Tagliaferri). 
The technical information about the observation is summarized in 
Table \ref{obslog}. \\
Calibrated and concatenated event lists were produced from 
the Observation Data Files (ODF) 
with {\verb xmmsas } v. 6.5.0, following
standard procedures  and with the most recent
calibration files available at the time of the analysis.  Due to a
bug of the {\verb omfchain } pipeline (SSC-SPR-3499)
the data from the Optical
Monitor (OM; Mason et al. 2001) were processed with {\verb xmmsas }
v. 6.1.0. \\
Pile-up effects in the EPIC PN (Str\"uder et al. 2001) and EPIC 
MOS (Turner et al. 2001)
data were checked with the task
{\verb epatplot }  and found to be 
acceptable (observed-to-predicted single and double event ratios $\sim 1$). 
We restricted the analysis of PN data in Timing Mode
to  the energy range over which 
the observed and predicted single and double event fractions
were found to agree well, i.e. between 0.5--10.0 keV.
For consistency, we used the same energy range for the analysis of the
MOS data, excluding as well any remaining  (cross-)calibration uncertainties 
below $\sim 0.3$ keV (Kirsch 2006). \\
As the observation was affected by soft-proton flares,
only Good Time Intervals (GTI)               
were selected for the analysis after the
inspection of  
the background light curves of the different instruments. The resulting
effective exposures are given in Table \ref{obslog}. One should notice that,
due to the use of the thick filter, the MOS 2 camera is less sensitive to
background flares and therefore the GTI selection yields a much         longer
effective exposure than for the PN and MOS1. The way this time selection
 affects the spectral analysis
is discussed in Sect. \ref{spectral}.\\
We extracted source counts for the PN 
from a central strip with $28\le {\rm RAWX}\le
47$ and background counts from two strips
at the sides of the CCD with  $2\le {\rm RAWX}\le 11$ and $54\le {\rm RAWX}\le 63$.  
Source counts for MOS1 and MOS2 were extracted from 
circular regions centered on the source with radii of 27  and 60 arcsec,
respectively, where the smaller radius for MOS1 is due to the use of
the Small Window Mode.
Background counts have been extracted from circles of the same size and
located on the same CCD as the
source regions. In the case of MOS1 in Small Window
Mode care was taken to choose a background region least affected
by the source counts. Only single and double events (PN) and
single-to-quadruple events (MOS) with quality FLAG=0 were used in the
analysis.\\
PN and MOS spectra were created with {\verb evselect } and grouped with 
{\verb grppha } in order to have at least 30 counts in each energy
bin for the use of the $\chi ^{2}$ statistics.  
Redistribution matrices and ancillary response files were produced with
{\verb rmfgen } and {\verb arfgen }, respectively.\\
RGS (den Herder et al. 2001) spectra and response
matrices have been created with {\verb rgsproc }. We have considered only the 
0.8-2.0 keV band, excluding 
lower energies where the RGS show a 30-40\% flux loss
with respect to the PN and MOS cameras (Stuhlinger et al. 2006).\\
The spectral analysis was performed with XSPEC v. 11.3.1., whereas the timing
analysis was carried out 
with FTOOLS v. 5.3.1. and our own routines. \\ 

Three other short observations (Obs. IDs 0012850101, 0012850601, 0012850701, 
PI J. Mulchaey),
in which \s5 was in the field of view, were found in the XMM archive. 
They were performed on April
13, 2001, September 19, 2001 and March 26, 2002, respectively.   
The target of these observations was the cluster 1WGA J0720.8+7108 at z=0.23,
whereas \s5 was lying at $\sim 12$ arcmin from the boresight. The PN and MOS
cameras were operated in Full Frame Mode with a thin filter, except for 
obs. ID  0012850601 for which no PN data were taken.
However, the source counts from these three observations
are much lower than that from the 2004 
data, both due to the shorter exposures and to the stronger
vignetting. Furthermore, sources
observed at off-axis angles larger than $\sim 10$ arcmin suffer from large
calibration and vignetting correction
uncertainties (Kirsch 2006). 
For the above
reasons we do not report on the detailed analysis
of these data; however, for completeness, in Sect. \ref{spectral} we quote the
fluxes of the source at 
the three different
epochs, using the results of the spectral analysis from the 2004 observation.\\
In the following, a cosmology  with $H_{\rm 0}=70$ km s$^{-1}$ Mpc$^{-1}$, $\Omega _{\rm M}=0.27$ and $\Omega
_{\rm \Lambda }=0.73$ is assumed.

\begin{table*}
\begin{center}
\small
\tabcolsep1ex
\caption{\label{obslog} Summary of the XMM-{\it Newton} observation of \s5
  during April 4--5, 2004.}
\begin{tabular}{cccccc}
\noalign{\smallskip} \hline \noalign{\smallskip}
\hline
\multicolumn{6}{c}{EPIC+RGS} \\
\noalign{\smallskip} \hline \noalign{\smallskip}
\multicolumn{1}{c}{Instrument} & \multicolumn{1}{c}{Mode} & 
\multicolumn{1}{c}{Filter} & \multicolumn{1}{c}{Energy band (keV)} &
\multicolumn{1}{c}{Effective exposure$^{*}$ (s)} & \multicolumn{1}{c}{Net
  count rate (cts/s)}\\
\noalign{\smallskip} \hline \noalign{\smallskip}
PN & Timing & Thin & 0.5-10.0 & 48491 & 4.42 \\
MOS1 & Small Window & Thin & 0.5--10.0 & 46808 & 1.59 \\
MOS2 & Full Frame & Thick & 0.5--10.0 & 54144 & 1.38 \\
RGS1 & Spectroscopy & - & 0.8--2.0 & 37025 & 0.22 (order 1) \\
     &              &   &         &       & 0.04 (order 2) \\
RGS2 & Spectroscopy & - & 0.8--2.0 & 35672 & 0.23 (order 1) \\
     &              &   &         &       & 0.03 (order 2) \\
\noalign{\smallskip} \hline \noalign{\smallskip}
\hline
\multicolumn{6}{c}{Optical Monitor} \\
\noalign{\smallskip} \hline \noalign{\smallskip}
\multicolumn{1}{c}{Mode} & \multicolumn{1}{c}{Filter} &
\multicolumn{1}{c}{Wavelength (nm)} & \multicolumn{1}{c}{N. exposures} &
\multicolumn{2}{c}{Total effective exposure (s)} \\
\noalign{\smallskip} \hline \noalign{\smallskip}
Imaging+Fast & V & 543 & 8 & \multicolumn{2}{c}{9600} \\
Imaging+Fast & U & 344 & 10 & \multicolumn{2}{c}{12000} \\
Imaging+Fast & UVW1 & 291 & 5 & \multicolumn{2}{c}{6000} \\
Imaging+Fast & UVM2 & 231 & 5 & \multicolumn{2}{c}{6000} \\
Imaging      & UVW1 & 291 & 5 & \multicolumn{2}{c}{4000} \\
Imaging      & UVM2 & 231 & 5 & \multicolumn{2}{c}{4000} \\
\noalign{\smallskip}\hline
\end{tabular}
\medskip

\end{center}
* After selection of Good Time Intervals.\\

\end{table*}

         %__________________________________________________________________

\section{Spectral analysis}
\label{spectral}

\begin{figure}
\psfig{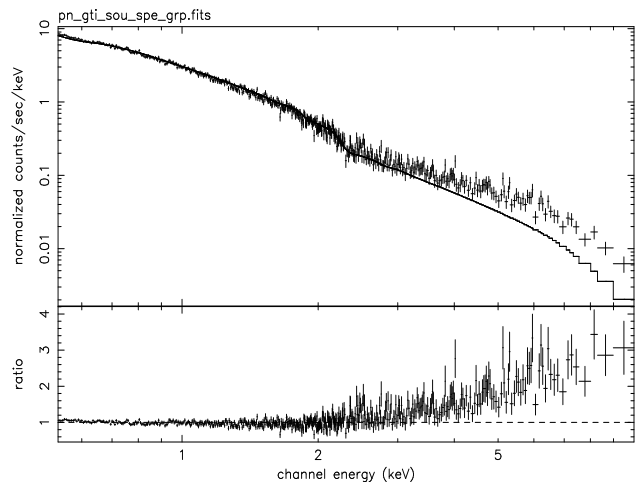}
\caption[]{Power law fit to the PN data (upper panel) with data-to-model
  ratios (bottom panel), clearly showing the upturn of the spectrum with
increasing energies. Similar residuals are obtained for the MOS data.}
\label{powl}
\end{figure}

\begin{figure}
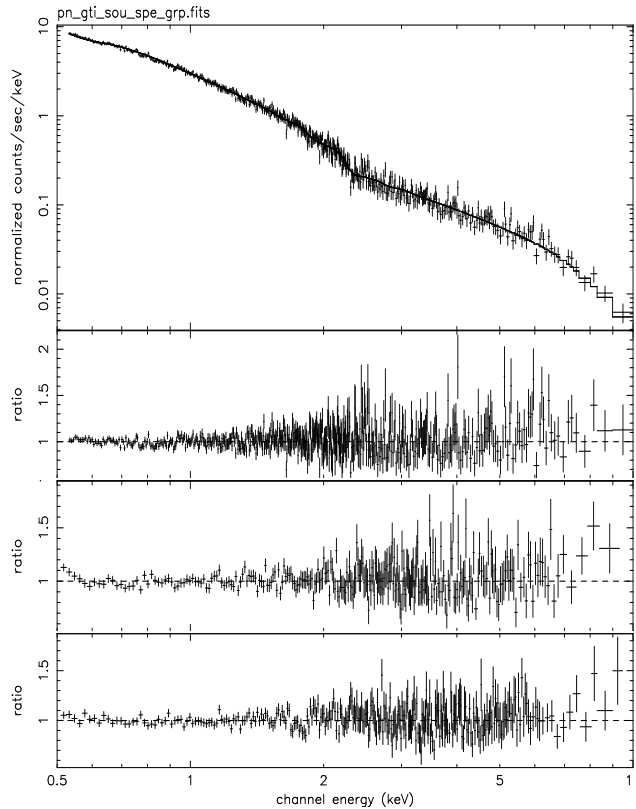

\psfig{figure=5317fg2a.ps,height=6.3truecm,width=8.3truecm,angle=-90,%
 bbllx=515.5pt,bblly=43pt,bburx=97pt,bbury=701pt,clip=}
\psfig{figure=5317fg2b.ps,height=2.0truecm,width=8.3truecm,angle=-90,%
 bbllx=515.5pt,bblly=43pt,bburx=385pt,bbury=701pt,clip=}
\psfig{figure=5317fg2c.ps,height=2.3truecm,width=8.3truecm,angle=-90,%
 bbllx=554pt,bblly=43pt,bburx=385pt,bbury=701pt,clip=}
\caption[]{Broken power law fit to the PN data (upper panel) with data-to-model
  ratios (second panel). Also shown are the ratios from the broken power law
  fits  for MOS1
  (third panel) and MOS2 (fourth panel).}
\label{bkntot}
\end{figure}

We started by  carrying out the spectral analysis for the entire
observation, 
without combining the data from different instruments.  We
restricted the analysis of PN and MOS data to the 0.5--10.0 keV band as 
explained in
Sect. \ref{dataproc}. In the spectral fits we used the Galactic neutral
hydrogen  column density measured towards the source by Murphy et
al. (1996), corresponding to  $N_{\rm H}=3.05\times 10^{20}$ cm$^{-2}$. 
All quoted errors are 90\% confidence ($\Delta \chi ^{2}=2.706$ for one 
interesting parameter) unless otherwise stated.  The results of the spectral
fits are summarized in Table \ref{fittotal}. We remark that
 the detailed numerical values for the three detectors may differ, 
because of the use of different GTIs 
combined with the strong spectral variability of the source (see later on),
but 
the physical models which describe the data should be the same. \\   
We first attempted to fit the data with the simplest model, a power law plus
Galactic absorption. The fit is clearly not acceptable, with large positive
residuals  above $\sim 3$
keV, indicating a flattening of the spectrum at high energies (see
Fig. \ref{powl}). Letting
$N_{\rm H}$ free to vary does not improve the fit significantly and yields an
absorption much lower than the Galactic value. \\
A good fit is obtained using a
broken power law plus Galactic absorption (see Fig. \ref{bkntot}). 
No need for extra absorption is
found when $N_{\rm H}$ is left as a free parameter. 
From the  broken power law model we obtain PN unabsorbed fluxes in the total
and hard bands of   
$F_{\rm 0.5-10~keV}=1.01^{\rm +0.01}_{\rm -0.02}\times 10^{-11}$ \ergs and
$F_{\rm 2-10~keV}=3.83^{\rm +0.12}_{\rm -0.14}\times 10^{-12}$ \ergs,
 respectively. 
These fluxes correspond to lower limits on the rest frame luminosities  of 
$L_{\rm 0.5-10~keV}\ge 3.42{\rm \pm0.03}\times 10^{45}$ \erglu and
$L_{\rm 2-10~keV}\ge 1.15^{\rm +0.03}_{\rm -0.02}\times 10^{45}$ \erglu 
  for $z\ge 0.3$.\\
A broken power law represents a good and simple parameterization of the X-ray
spectrum of \s5 in the XMM band; however, other spectral models can in
principle fit the data equally well (see Table
\ref{fittotal}). A double power law with
Galactic absorption provides an equally good fit.  Similarly, a power law
plus the addition of a black body component is also an acceptable fit, whereas
a power law plus a thermal bremsstrahlung component is an adequate fit for PN
and MOS2 data, but can be accepted only at 1\% significance 
level for MOS1. \\
We further attempted to fit the data with a logarithmic parabola of the form 
$E^{-\Gamma +\beta \log E}$. Although this is also a viable model, 
when compared with the broken power law or the double power law, 
it yields a worse fit. \\
Despite the fact that all the above-mentioned models, except for the single power law,
constitute valid
representations of the data, in agreement with previous X-ray studies and with
the 
interpretation of the SED of \s5, we adopt the broken and double power laws as 
our
favorite models.  
Both models can be 
interpreted easily (see
Sect. \ref{background}), 
 in terms of a mixed contribution of the
hard  energy tail of the synchrotron
emission (steeper, soft component) and the rising side of the IC
 bump (flatter, hard component). The logarithmic parabola 
can also be interpreted in this way, with the difference that it represents a
continuously curved spectrum instead of the combination of two straight power
laws. However, we discard it, as it provides a worse fit to
the data.\\
Conversely, a  model comprising a power law and a black body component
cannot be easily reconciled with any physical scenario for the source. The
black body emission might be attributed to the accretion disk; however, we do
not expect this component to contribute significantly in a BL Lac such as \s5,
where the jet emission is dominant in all bands and 
basically no sign of thermal emission has ever been observed at any other
wavelength. Furthermore, the temperature resulting
from the fit is rather high and therefore the black body component 
appears to merely provide a parameterization of
the high-energy part of the X-ray spectrum, without any real physical 
meaning.\\
A thermal bremsstrahlung component might originate from the host galaxy of the
source or from a cluster around it, or close to it. \
The PN unabsorbed 0.2--3.5 keV flux of the bremsstrahlung component
is  $4.92^{+0.27}_{-0.14}\times 10^{-12}$ \ergs, corresponding to a lower
limit on  the
luminosity for $z>0.3$ of $1.87^{+0.11}_{-0.06}\times 10^{45}$ \erglu . 
This luminosity largely exceeds
typical values 
for massive elliptical galaxies (Fabbiano 1989)
and therefore, similarly to Cappi
et al. (1994), we conclude that thermal emission from the host
galaxy is unlikely.
The cluster 1WGA J0720.8+7108 (see Sect. \ref{dataproc}) could be able in
principle to produce such a
high luminosity; however, at the location of \s5, about 12 arcmin  from the center 
of the cluster, it is implausible that thermal emission can contaminate
significantly the flux from the BL Lac. From a time-resolved spectral analysis
(the intervals used are those defined in Fig. \ref{totlight}, see
Sect. \ref{timespectral}),
the 0.2--3.5 keV
flux of the
bremsstrahlung component resulted to vary significantly 
($\Delta F \approx 44$\%), arguing further against a host galaxy
or cluster origin. \\
In conclusion, in the following discussions we will consider  only the broken 
and double power law models and we will make the implicit association between
  the steeper (flatter) power law with the synchrotron (IC) component.
We can thus 
see that, e.g. considering the PN, the synchrotron emission contributes to
$\sim 65$\% of the flux in the 0.5--10.0 keV band, whereas the IC constitutes a
smaller, but not negligible, $\sim 35$\%.\\

\begin{table*}
\begin{center}
\small
\tabcolsep1ex
\caption{\label{fittotal} Results of the spectral analysis for the entire
  XMM observation of \s5, on April 4--5, 2004. The neutral hydrogen column
  density ($N_{\rm H}$) has been fixed to the
  Galactic value in all fits.}
\begin{tabular}{cccccccc}
\noalign{\smallskip} \hline \noalign{\smallskip}
\hline
\multicolumn{1}{c}{Model} & \multicolumn{1}{c}{Instrument} & 
\multicolumn{1}{c}{$\Gamma _{1}$} &
\multicolumn{1}{c}{$E_{\rm br}$ (keV)/$\beta ^{*}$} & 
\multicolumn{1}{c}{$\Gamma _{2}/kT$ (keV)} &
\multicolumn{1}{c}{$\chi ^{2}_{\rm red}/{\rm dof}$} &
\multicolumn{1}{c}{Prob.} & \multicolumn{1}{c}{$F_{\rm 2-10~keV}^{\dagger}$}
\\ 

\noalign{\smallskip} \hline \noalign{\smallskip}
\vspace{0.1cm}
Power law & PN & 2.73$\pm $0.01 & - & - & 1.38/1451 & 9.30$\times 10^{-21}$ &
2.47$^{+0.04}_{-0.05}$\\
\vspace{0.1cm}
          & MOS1 & 2.53$\pm $0.01 & - & - & 1.90/281 & 4.90$\times 10^{-18}$ &
3.62$^{+0.08}_{-0.07}$\\
\vspace{0.1cm}
          & MOS2 & 2.50$\pm $0.01 & - & - & 1.66/306 & 2.47$\times 10^{-12}$ &
3.84$^{+0.07}_{-0.06}$\\
\vspace{0.1cm}
Broken power law & PN & 2.83$\pm $0.01 & 1.91$^{+0.10}_{-0.08}$ &
   2.06$^{+0.05}_{-0.06}$ & 0.982/1449 & 0.680 &
3.83$^{+0.11}_{-0.10}$\\
\vspace{0.1cm}
                 & MOS1 & 2.64$\pm $0.02 & 2.34$^{+0.25}_{-0.16}$ &
   2.02$^{+0.10}_{-0.07}$ & 1.070/279 & 0.200 &
4.43$\pm$0.13\\
\vspace{0.1cm}
                 & MOS2 & 2.62$\pm $0.02 & 2.27$^{+0.36}_{-0.20}$ & 
   2.11$^{+0.06}_{-0.11}$ & 0.957/304 & 0.695 &
4.46$\pm$0.09\\
\vspace{0.1cm}
Double power law & PN & 3.01$\pm $0.05 & - & 1.22$^{+0.15}_{-0.22}$ 
   & 0.982/1449 & 0.686 &
3.95$^{+0.45}_{-1.16}$\\
\vspace{0.1cm}
                 & MOS1 & 2.72$^{+0.08}_{-0.04}$ & - & 0.70$^{+0.45}_{-0.30}$
   & 1.038/279 & 0.318 &
4.69$^{+0.91}_{-1.39}$\\
\vspace{0.1cm}
                 & MOS2 & 2.74$^{+0.10}_{-0.06}$ & - & 1.13$^{+0.35}_{-0.36}$
   & 0.919/304 & 0.841 &
4.62$^{+0.79}_{-1.44}$\\
\vspace{0.1cm}
Power law + black body & PN & 2.87$\pm $0.02 & - & 1.71$^{+0.14}_{-0.12}$
   & 0.978/1449 & 0.721 &
3.45$^{+0.27}_{-0.29}$\\
\vspace{0.1cm}
                       & MOS1 & 2.67$\pm $0.02 & - & 2.41$^{+0.56}_{-0.33}$
   & 1.051/279 & 0.268 &
4.59$^{+0.45}_{-0.70}$\\
\vspace{0.1cm}
                       & MOS2 & 2.65$\pm $0.03 & - & 2.05$^{+0.29}_{-0.19}$
   & 0.922/304 & 0.830 &
4.45$^{+0.31}_{-0.44}$\\
\vspace{0.1cm}
Power law + bremss. & PN & 2.12$^{+0.04}_{-0.06}$ & - & 0.31$^{+0.02}_{-0.01}$
   & 1.018/1449 & 0.307 &
3.75$^{+0.09}_{-0.10}$\\
\vspace{0.1cm}
                    & MOS1 & 2.11$\pm $0.07 & - & 0.38$\pm $0.03 
   & 1.201/279  & 0.012 &
4.37$^{+0.15}_{-0.22}$\\
\vspace{0.1cm}
                    & MOS2 & 2.17$^{+0.05}_{-0.07}$ & - & 
   0.37$^{+0.04}_{-0.03}$ & 1.014/304 & 0.422 &
4.44$^{+0.14}_{-0.16}$\\
\vspace{0.1cm}
Logarithmic parabola & PN & 2.79$\pm $0.01 & 0.57$\pm $0.03 & - & 
   1.008/1450 & 0.413 &
3.76$^{+0.11}_{-0.09}$\\
\vspace{0.1cm}
                     & MOS1 & 2.66$\pm $0.02 & 0.44$\pm $0.04 & - & 
   1.108/280 & 0.104 &
4.44$^{+0.11}_{-0.12}$\\
\vspace{0.1cm}
                     & MOS2 & 2.64$\pm $0.02 & 0.38$\pm $0.04 & - &
   0.955/305 & 0.703 &
4.50$^{+0.11}_{-0.10}$\\

\noalign{\smallskip}\hline
\end{tabular}
\medskip

\end{center}
* $\beta $ is the ``curvature parameter'' in the logarithmic parabola model
  (see text for the analytical formula). \\
$\dagger$ in units of $10^{-12}$ \ergs

\end{table*}

A matter of concern are the significant discrepancies observed between the PN
 and MOS best fit parameters (see Table \ref{fittotal}). In fact, it is impossible 
to fit the PN and MOS data jointly 
(e.g. a broken power law yields $\chi ^{2}_{\rm red}/{\rm dof}=1.110/2038$, prob.=$3.37\times 10^{-4}$).  Only when 
the parameters for the three cameras  are left free to
vary independently, except for the break energy, the fit becomes acceptable
 ($\chi ^{2}/{\rm dof}=0.998/2034$,
prob.=0.516). However, the photon indices still differ by $\Delta \Gamma \sim 0.2$
and the fluxes by $\sim $15--16\%. Although PN/MOS differences
are widely known, the flux discrepancies that we observe 
appear to be larger than
what is reported in the XMM documentation (Stuhlinger et al. 2006).  
On the other hand, to our knowledge, investigations
 on  PN/MOS cross-calibration uncertainties have so far not taken into
 account 
the Timing Mode,
 and thus we cannot compare directly our results with 
previous studies.\\
We checked whether 
the spectral variability of \s5,
combined with the different pattern of GTI selection 
for the three instruments 
(see Sect. \ref{dataproc}), could explain (at least part of) the observed
disagreement (an extensive discussion of the
spectral variability of \s5 will be given in Sect. \ref{timespectral}).  
To this purpose, we extracted data from a time
interval, common to PN and MOS, characterized by 
similar GTIs created by the standard processing pipeline, which include
periods of normal functioning of the instruments. The initial $\sim 19$ ks of
the observation fulfilled this criterion and were selected for the following
analysis.  
We then applied the user-defined GTIs, based on the background light curve
 of the PN camera,
also to MOS1 and MOS2 (we notice that no case exists in which a background
 flare is registered by 
 MOS1 or MOS2 but not by the PN; therefore we do not risk to include
 periods of high background in the MOS data). 
In this way we are sure to sample similar time
intervals for the three cameras, with effective exposures differing by no more
than $\sim 500 $ s. 
A joint broken power law fit to the PN and MOS spectra extracted from the
 above time
interval was
found to 
be acceptable ($\Gamma _{\rm 1}=2.85\pm 0.02$, $E_{\rm break}=1.81^{\rm
  +0.11}_{\rm -0.10}$ keV, $\Gamma _{\rm 2}=2.17\pm 0.05$, $\chi
^{2}_{\rm red}/{\rm dof}=0.908/631$, prob.=0.953), contrary to the case of the whole
observation. The flux differences are significantly
diminished to $\sim 7-10$\%, a value closer to, although still higher than
 those normally reported for the
Imaging modes (Stuhlinger et al. 2006).
 We conclude that
the spectral variability of the source gives a considerable contribution to 
the discrepancies
between PN and MOS, although it cannot account entirely for it.
Further investigations of the remaining discrepancies are beyond the scope
 of this paper; however,
 we remark that the errors on all the quoted fluxes do not include the 
 $\sim 7-10$\% uncertainty.\\

Using the PN and MOS data (separately) for the whole observation, 
 we further estimated an upper limit on
any possible intrinsic absorption in the source. To
 this purpose, we used
the best fit broken power law model with the  
addition of a second redshifted component (model {\verb zwabs } in 
{\verb xspec }) to the Galactic neutral
hydrogen column density. We tried
a range of values for the redshift and found that, up to z=2, 
values of $N_{\rm H,int}\ga 2\times 10^{20}$ cm$^{-2}$ are not compatible with
 the
data.\\

Given the importance that any detection of a line would
 have for the
 determination of the redshift of \s5,
we looked in more detail for emission and absorption features in the
EPIC spectra.
By using the broken power law as the base model, we added to it
a Gaussian emission or absorption line.  
We stepped the initial value of
the line energy over the entire spectrum through a grid of 0.5 keV bin
size and we attempted a fit for each grid point, either 
with both the energy and the width of the line $\sigma $ left free to vary, or 
with $\sigma $ fixed to 0.01 keV (i.e. a narrow line,
below the energy
resolution  of
the instruments). If the fit did not converge to any line energy, the latter
was fixed to the grid point and an upper limit
to either the equivalent width (EW) or to the optical depth was determined. 
With this procedure no line was found to be statistically significant: 
the fits did not improve after the addition of a
line at any energy within the
 0.5--10.0 keV band and no line is detected at the same energy 
in the spectra of the three
cameras. In  all
cases the minimum EWs and optical depths compatible with the data are equal to
zero. Upper limits on EWs for energies up to $\sim 5.8$ keV (corresponding to
 a neutral Fe K$\alpha $ line at $z\ge0.1$) are of the order of $\sim 100$ eV. 
We also examined the RGS data, but no
line could be found.
As the RGS did not provide better or additional information
with respect to the PN and
MOS data, we do not consider them further. \\

Finally, we estimated the flux for the three observations in 2001--2002 by
fitting a broken power law with all the parameters fixed to the 2004 
best fit values for each instrument 
(see Table \ref{fittotal}). Only the normalizations were left free to vary.
The fits were acceptable for the 2001 observations
and thus we have no reason to believe that the spectrum of the source was
significantly different at that time.
The comparison of the 0.5--10.0 keV MOS1 (the
PN was off during the September exposure)
unabsorbed fluxes of the 2001 observations ($6.68^{+0.85}_{-0.45}\times
10^{-12}$ \ergs and  $9.13^{+0.80}_{-0.62}\times 10^{-12}$ \ergs) with that in
2004 ($F_{\rm 0.5-10.0~keV}=1.13\pm 0.01\times 10^{-11}$ \ergs) shows that the
source was about a factor of two fainter in the earlier epoch, but with a
similar flux level in the second epoch.
On the other hand,
a broken power law fit to the PN spectrum of the 2002 observation is 
acceptable only at 1\% significance level and it fails when
applied to the MOS data. This might be an indication that significant spectral
variability took place between 2002 and 2004. 
However, the MOS flux in 2002, which appears to be
 rather insensitive to
the parameter values, is about three times larger than the PN flux,
casting doubts on the reliability of these data. The PN flux 
($F_{\rm 0.5-10.0~keV}=8.55^{+0.35}_{-0.38}\times 10^{-12}$ \ergs)  is more in
line with the values that have been observed so far for \s5.

%___________________________________________________________________

\section{Temporal analysis}
\label{timing}

\begin{figure}
\psfig{figure=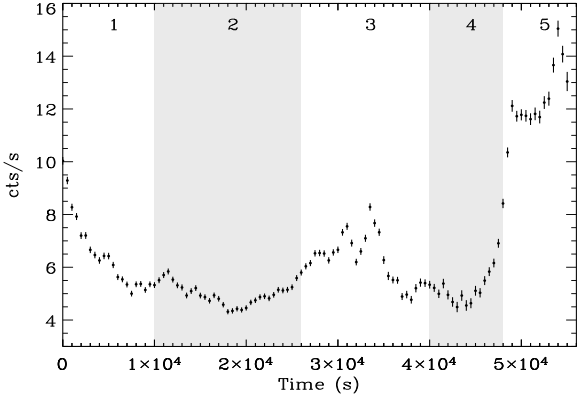,height=5.3truecm,width=8.3truecm,angle=90,%
 bbllx=63pt,bblly=712pt,bburx=454pt,bbury=170pt,clip=}
\caption[]{Combined PN+MOS light curve in the 0.5-10.0 keV band. The bin
  size is 500 s and the time is counted from the beginning of the
  observation (JD=2453099.983). 
The figure also shows the time intervals used for the
  time-resolved  spectral analysis (see Sect. \ref{timespectral}).}
\label{totlight}
\end{figure}

As a first step, we produced 
 the 0.5--10.0 keV band background-subtracted source light curves for 
the three 
EPIC cameras separately,
with a time bin size of 500 s. 
Photons were extracted from a time interval common to all
the instruments, as the three cameras start to register events at slightly
different times.  The resulting effective exposures for the timing analysis
are $\sim $55 ks. 
According to a Kolmogorov-Smirnov test, the ratios of the light curves from
 any
pairs of instruments
resulted to be consistent with being constant
at 5\% significance level. The light curves obtained from the three cameras, with different operating modes, filters and backgrounds,  
are thus consistent with each other and are not affected by periods of high 
background, especially important towards the end of the observation. 
Therefore, we summed the PN and MOS light curves up to increase the number
statistics in each bin. The combined 0.5--10.0 keV light curve is shown in
Fig. \ref{totlight}. 
Strong variability is
clearly detected, with the shortest observed doubling time of $\sim 2.5$ ks,
 corresponding to the steep rise of the flux around $t\sim 5\times 10^{4}$ s.  
The maximum-to-minimum count rate
 ratio is $\ga 3$.
The mean count rate is $6.59\pm 0.01$ cts/s, with an
observed variance (i.e. including
measurement noise) of 6.14 (cts/s)$^{2}$. \\
The beginning of the observation caught the source in a phase of decreasing
 flux. 
During the first $\sim 7-8$ ks the count rate
reduced  
by a factor of $\sim 1/2$. The decreasing trend continued
for another $\sim 10$ ks, with a slightly shallower slope and 
with a few small amplitude bursts.
Afterwards, the count rate started to increase again, at
first
more slowly for $\la 8$ ks, and then resulting in a burst characterized by  
two or, possibly, three peaks,  
lasting in total $\sim 10$ ks. 
This high state is followed by a
relatively quiescent period of $\sim  8$ ks, characterized by variations of
 smaller amplitude, preceding a big burst towards
the end of the observation. At the beginning of this flare
 the count rate increased by a factor of $\sim 3$  in
$\sim 4$ ks, settled itself on a flat level for $\sim 3$ ks, reached the
maximum value of the entire observation in $\sim 2$ ks and decayed again. 
The decreasing phase of this big flare was unfortunately not fully sampled.\\
A measure of the intrinsic source variability is given by the
fractional variability amplitude (FVA; Edelson et al. 2002; Vaughan et
al. 2003). In the 0.5--10.0 keV band we found 
${\rm FVA}=37\pm 2$\%. The uncertainty was calculated according to the formula
 given in Vaughan et
al. (2003) and is due to measurement errors only.

\begin{figure}
\psfig{figure=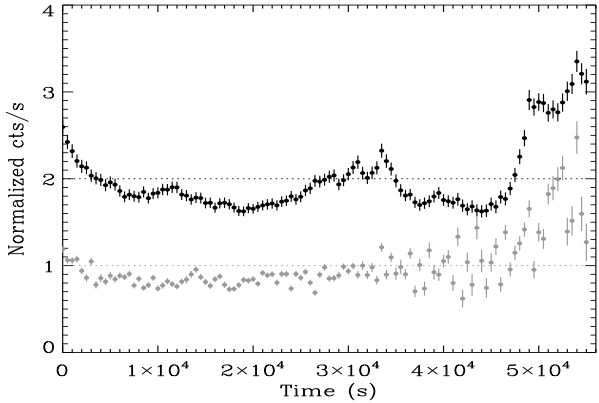,height=5.3truecm,width=8.0truecm,angle=90,%
 bbllx=67pt,bblly=689pt,bburx=453pt,bbury=250pt,clip=}
\caption[]{Combined PN+MOS light curves
  in the 0.5--0.75 keV (black) and 3--10 keV (gray) bands. The light curves 
  have been
  normalized to their mean count rates and shifted with respect to each other
  for clarity. The mean level for each normalized light curve is shown with
  a dotted line of corresponding color. The bin size
  is 500 s.}
\label{sublight}
\end{figure}

We checked 
whether and to what extent the fractional variability amplitude
depended on the energy band considered. 
We extracted source light curves in two different energy
bands:  
a soft one (0.5--0.75 keV) and a hard one
(3.0--10.0 keV). According to the double power law spectral model 
(see Sect. \ref{spectral}),
the contribution  of the synchrotron
(IC) component to the total flux is $\sim 89$\% ($\sim 11$\%) for the soft 
band and 
$\sim 29$\% ($\sim 71$\%) for the hard band. The soft and hard light curves,
normalized to their mean count rates, are shown in Fig. \ref{sublight}. 
The variability amplitude
appears more
pronounced in the soft band.
The FVA in the soft band
($40\pm 3$\%) is indeed higher than in the hard band
($27\pm 1$\%). 

\begin{figure}
\psfig{figure=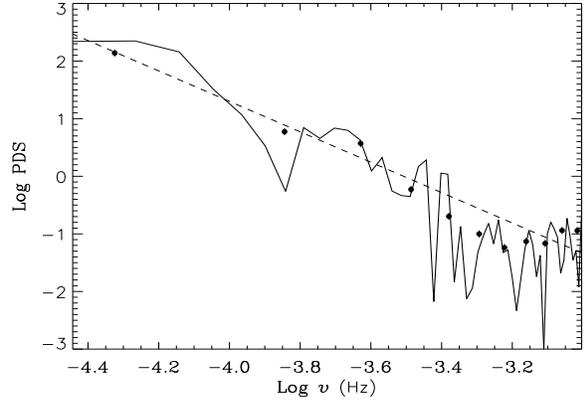,height=5.3truecm,width=8.3truecm,angle=90,%
 bbllx=66pt,bblly=717pt,bburx=454pt,bbury=170pt,clip=}
\caption[]{Power density spectrum of the combined PN+MOS light curve in the
  0.5--10.0  keV band. The bin
  size is 500 s. The dots show the re-binned PDS (see text). The dashed line indicates the
power law fit with slope of $2.64\pm0.36$.}
\label{totpow}
\end{figure}

In order to study the statistical characteristics of
the variability of the source, we calculated the power density spectrum 
(PDS; Priestley 1981) of the 0.5--10.0 keV light curve between $\sim 4\times
10^{-5}$ Hz and $10^{-3}$ Hz.  
The result is shown in Fig. \ref{totpow}. The PDS does not show any precise 
characteristic periodicity or time scale in this time range. 
 The PDS was re-binned according to the
 method  of Papadakis \& Lawrence (1993) and fitted with a broken power law.
 The break, above which the noise level is reached, was found at 
$\nu \sim 5\times 10^{-4}$ Hz.
 The fit gave a slope below the break of $2.64 \pm 0.36$, typical of a red
 noise process. The fastest variability time scale, before falling into noise, 
appears to be $\sim 2.5 $ ks, corresponding to the smallest doubling time
observed in the light curve.
The slope of the PDS was not found to depend significantly on the
energy band considered.

\begin{figure}
\psfig{figure=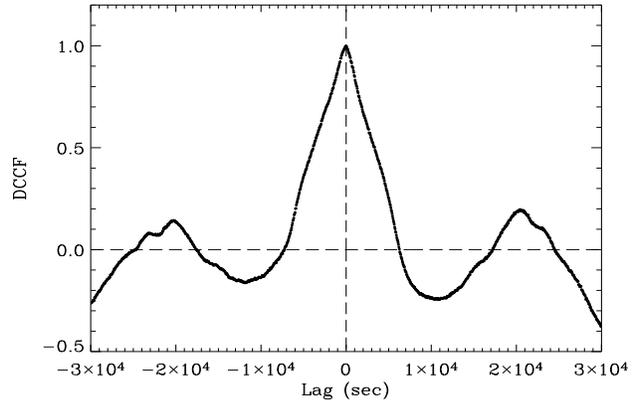,height=5.3truecm,width=8.3truecm,angle=90,%
 bbllx=66pt,bblly=715pt,bburx=453pt,bbury=170pt,clip=}
\caption[]{Discrete cross-correlation function for the soft (0.5--1.0 keV) 
and hard (1.0--10.0 keV) bands. 
The binning is 50 s.}
\label{ccfsh}
\end{figure}

To investigate the possible presence of time lags between 
light curves in different bands, we calculated the corresponding 
discrete cross-correlation
functions (DCCF; Edelson \& Krolik 1988). The bands used 
for this analysis were: 0.5--1.0 keV (soft), 1.0--10.0 keV (hard), 0.5--0.75 keV (A), 0.75--1.2
keV (B), 1.2--10.0 keV (C).
In order to be able to detect lags as small as a few hundred seconds,
we used a finer binning of 50 s.
We have determined 
the position of the peak ($\tau _{\rm peak}$) of each DCCF. 
The uncertainties on $\tau _{\rm peak}$ were estimated
through simulations (1000 runs), according to the ``bootstrap'' method 
described in
Peterson et al. (1998). These uncertainties account for errors due to 
measurement noise and
sampling.  
The results of the cross-correlation analysis are
given in Table \ref{cross}, and  the DCCF for the
soft and hard bands is shown in Fig. \ref{ccfsh} (similar functions are
obtained for the other bands). 
Negative lags mean that the soft band is lagging behind
the hard band (soft lags). \\ 
The distributions of  $\tau _{\rm peak}$ indicate
that lags $\ga 100$ s were not present between any of the bands, although 
smaller values cannot be excluded.

\begin{table}
\begin{center}
\small
\tabcolsep1ex
\caption{\label{cross}  Results of the cross-correlation analysis with
  a bin size of 50 s. The values of $\tau _{\rm peak}$
 are the medians of the respective
   distributions and the uncertainties represent the $\sim 68$\% limits
  (corresponding to $1\sigma $ for a Gaussian distribution). The bands are the
  following: 0.5--1.0 keV (soft), 1.0--10.0 keV (hard), 0.5--0.75 keV (A),
  0.75--1.2 keV (B), 1.2--10.0 keV (C).
}
\begin{tabular}{ccc|c}
\noalign{\smallskip} \hline \noalign{\smallskip}
\hline
\multicolumn{1}{c}{Bands} & \multicolumn{1}{c}{$\tau _{\rm peak}$ (s)} &
\multicolumn{1}{c}{CCF$_{\rm max}$} & 
\multicolumn{1}{c}{$\tau _{\rm peak,sim}$ (s)} \\
\noalign{\smallskip} \hline \noalign{\smallskip}
\vspace{0.1cm}
Soft vs. hard & -50 & 1.00$\pm $0.06 & 
$-50^{+125}_{-75}$ \\
\vspace{0.1cm}
A vs. B & 0 & 1.00$\pm $0.06 &  $0^{+125}_{-175}$ \\
\vspace{0.1cm}
B vs. C & +50 & 1.01$\pm $0.07 &  $-50^{+125}_{-125}$ \\
\vspace{0.1cm}
A vs. C & -50 & 0.99$\pm $0.06 &  $-50^{+125}_{-125}$ \\

\noalign{\smallskip}\hline
\end{tabular}
\medskip

\end{center}
\end{table}

%____________________________________________________________________

\section{Spectral variability}
\label{timespectral}

\begin{figure}
\psfig{figure=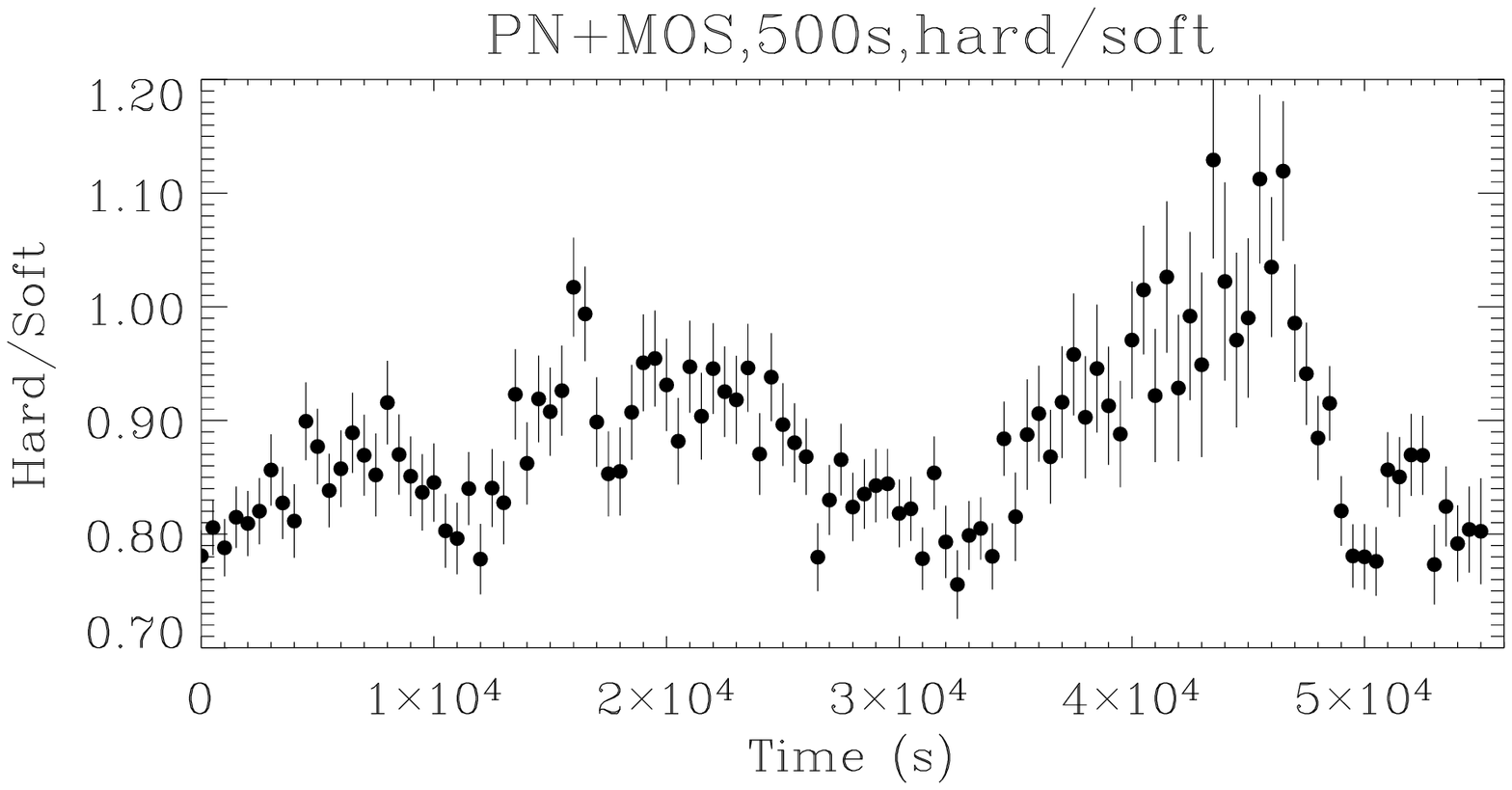,height=2.3truecm,width=8.3truecm,angle=90,%
 bbllx=112pt,bblly=712pt,bburx=310pt,bbury=195pt,clip=}
\psfig{figure=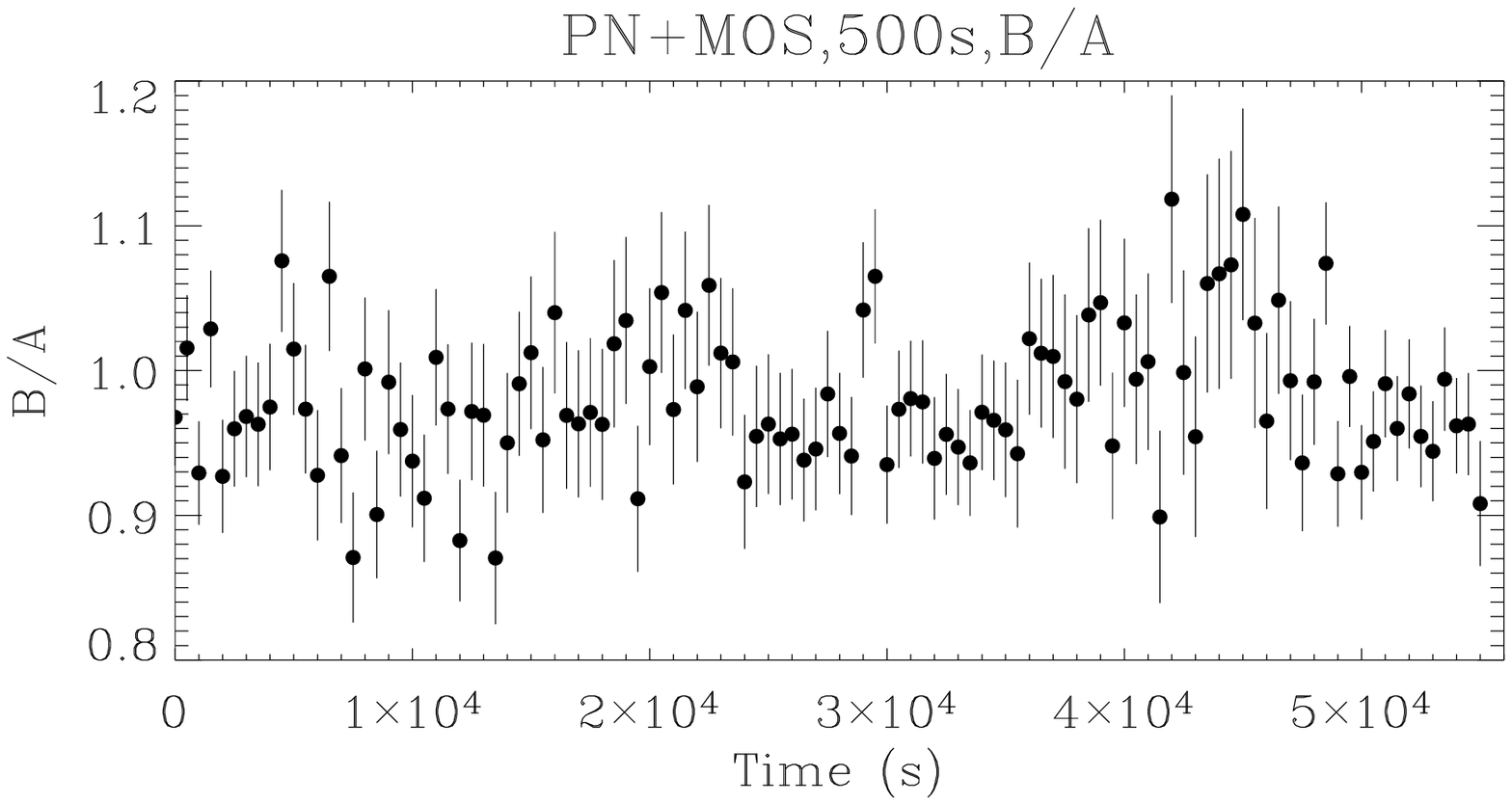,height=2.3truecm,width=8.3truecm,angle=90,%
 bbllx=112pt,bblly=712pt,bburx=310pt,bbury=195pt,clip=}
\psfig{figure=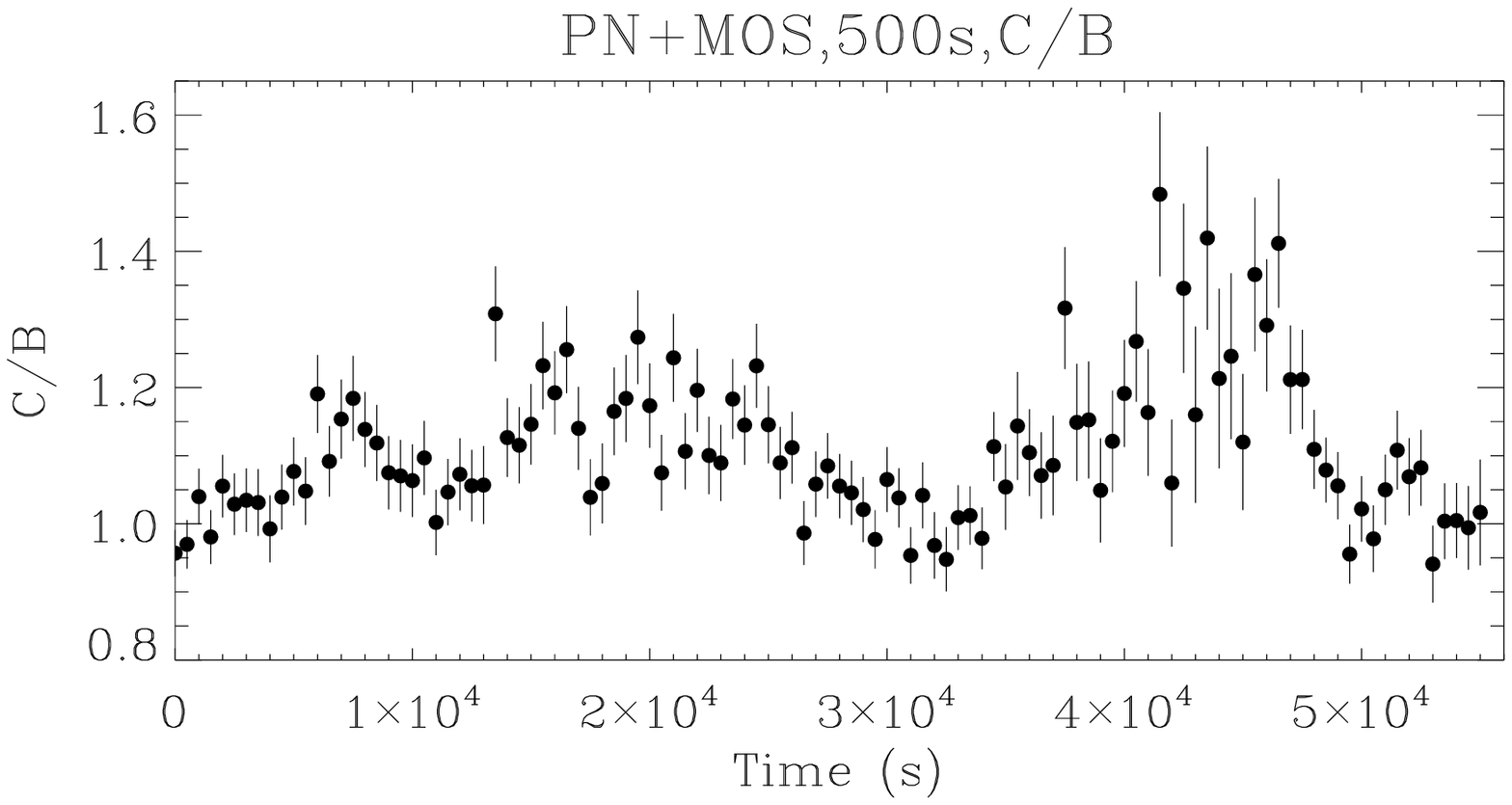,height=2.3truecm,width=8.3truecm,angle=90,%
 bbllx=112pt,bblly=712pt,bburx=310pt,bbury=195pt,clip=}
\psfig{figure=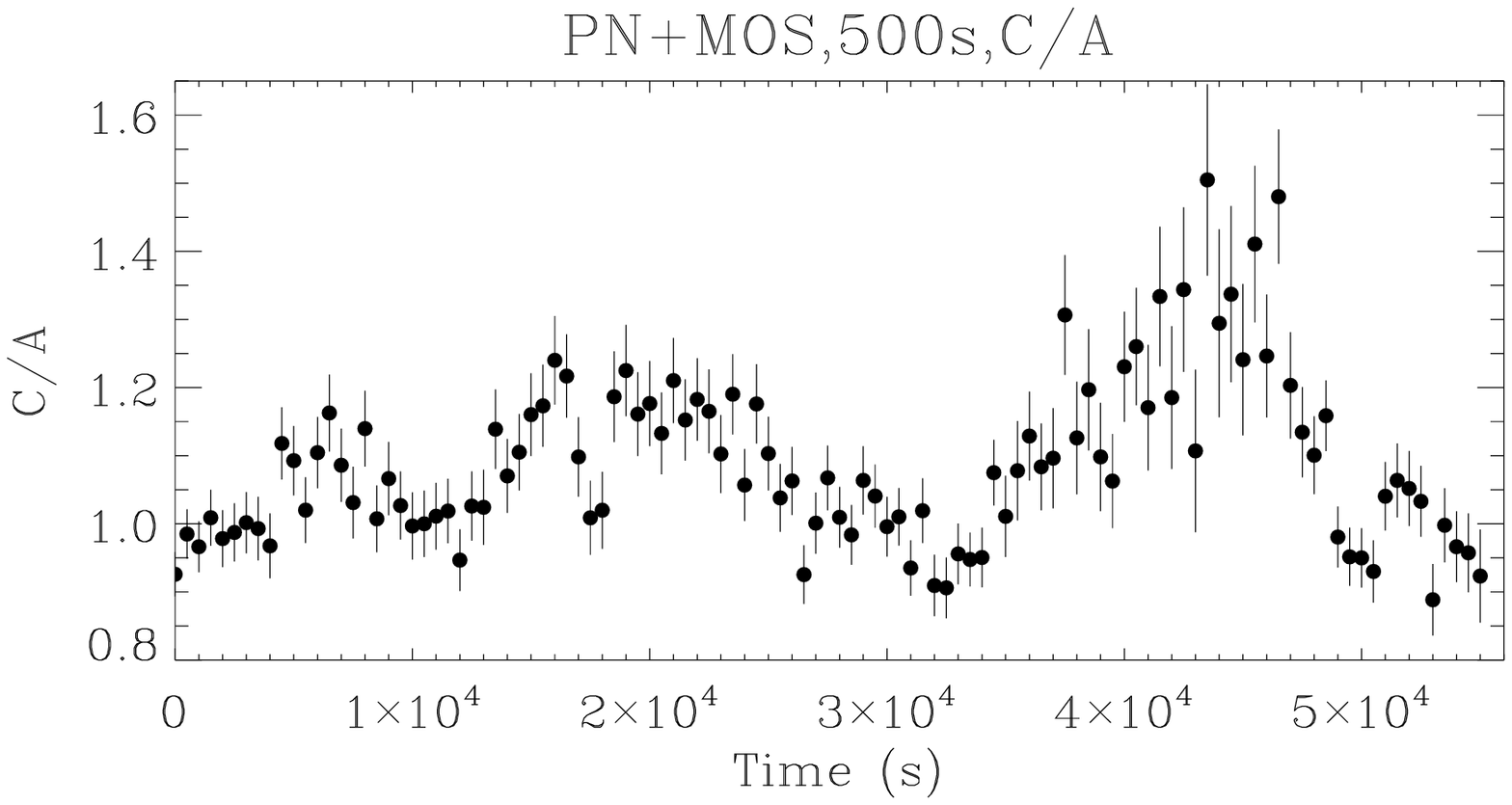,height=2.3truecm,width=8.3truecm,angle=90,%
 bbllx=112pt,bblly=712pt,bburx=310pt,bbury=195pt,clip=}
\psfig{figure=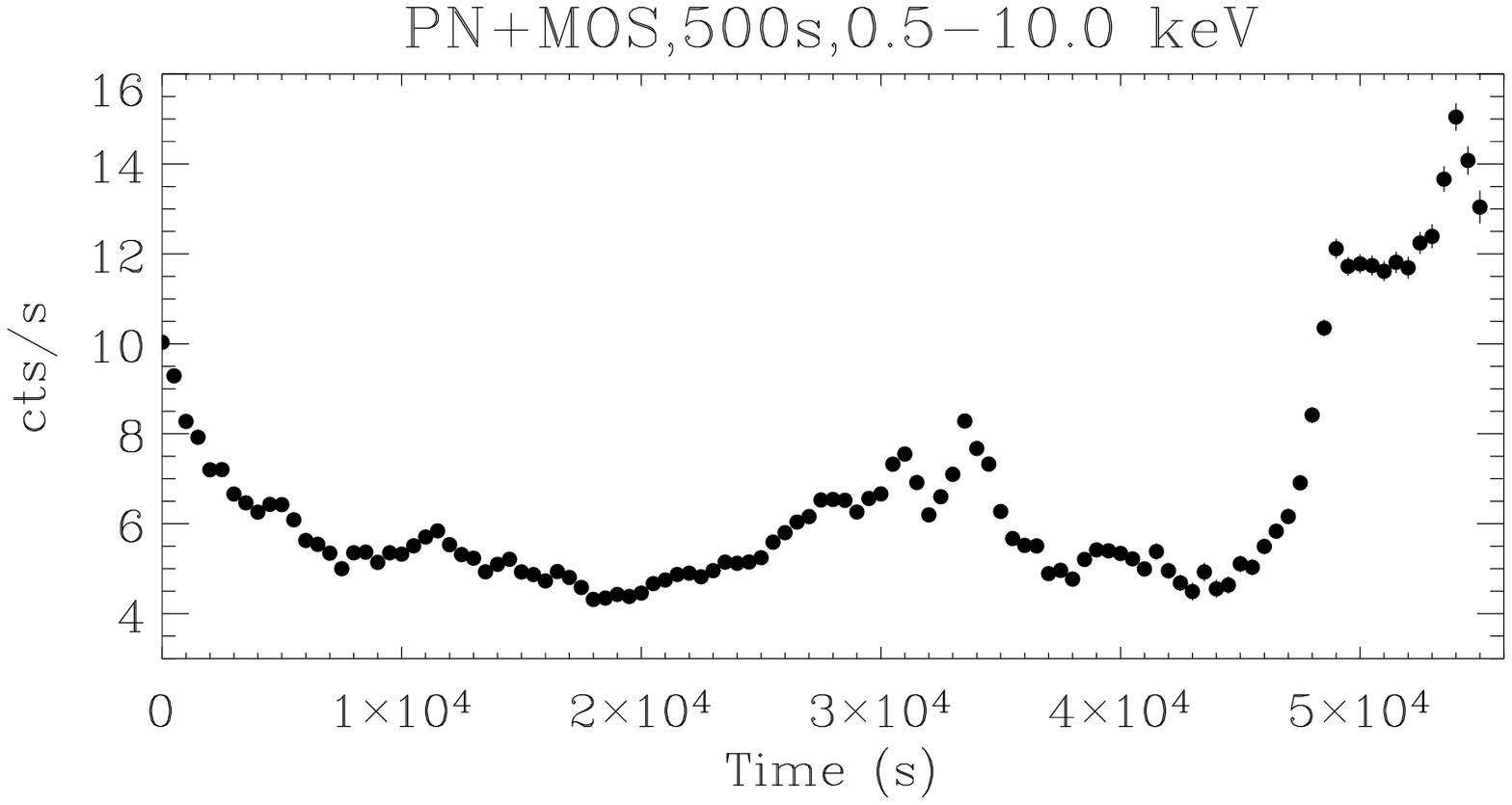,height=2.6truecm,width=8.3truecm,angle=90,%
 bbllx=67pt,bblly=712pt,bburx=310pt,bbury=195pt,clip=}
\caption[]{Hardness-ratio light curves for different bands: hard and soft 
(upper
  panel), A and B (second panel), B and C (third panel) and A and
  C  (fourth
  panel). The lower panel shows the 0.5--10.0 keV light curve for comparison.}
\label{hr}
\end{figure}

In this Section we investigate the
spectral variability of \s5 
and the relation with flux
variations. Hardness-ratio
light curves using the various sub-bands defined for the
  cross-correlation analysis in
Sect. \ref{timing} (see Table
  \ref{cross}) 
are shown in Fig. \ref{hr}. 
The hardness-ratios appear to be anticorrelated with respect
to the total count rate,
i.e. the spectrum softens when the
source brightens. This is different (but see discussion below)
from the usual
harder-when-brighter trend observed for HBL (see Sect. \ref{intro}).

\begin{figure}
\psfig{figure=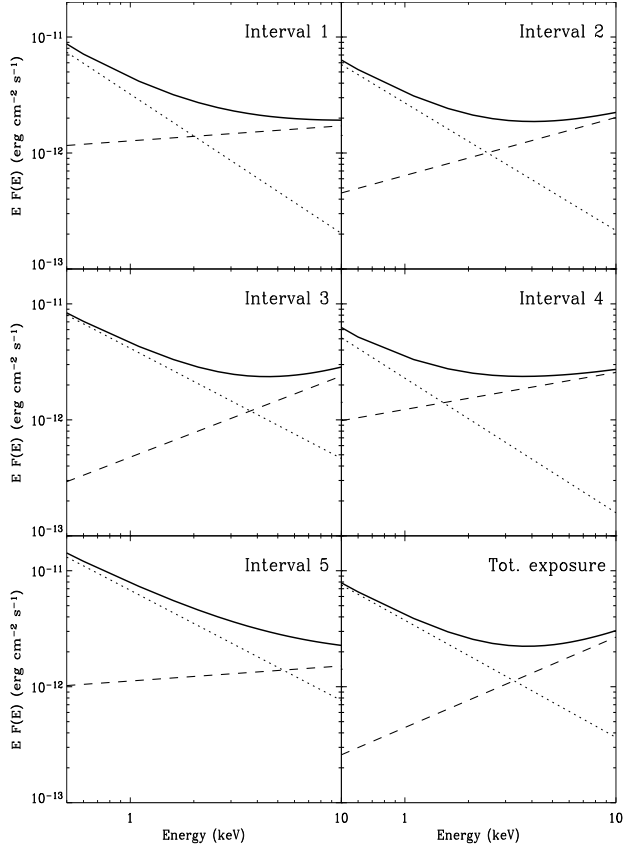,height=11.3truecm,width=8.3truecm,angle=90,%
 bbllx=376pt,bblly=737pt,bburx=865pt,bbury=323pt,clip=}
\caption[]{Spectral variations during the observation, derived in the case of
 the double power
  law model. The dotted and dashed lines represent the steep (synchrotron)
  and flat (IC) power law components, whereas the solid line is the sum of
  the two. The time intervals are defined in Fig. \ref{totlight}. The bottom
 right panel shows the SED for the total exposure. }
\label{specvar}
\end{figure}

In order to investigate the spectral variability in greater detail we divided
the observation into five intervals (see Fig. \ref{totlight}).
The subdivision shown in Fig. \ref{totlight} 
was the result of a compromise between
the need to have a reasonable number of time intervals, necessary
 to investigate the
evolution of the spectral parameters, and a sufficiently large
number of counts for the
spectral analysis in each interval.  
With  shorter 
time intervals it was impossible to 
constrain 
the spectral parameters significantly
and estimate the fluxes of
the  synchrotron and IC components separately.\\
We 
repeated the spectral analysis for each interval shown in Fig. \ref{totlight}, 
fitting the spectra
with a double power law with Galactic absorption.  All fits were acceptable
 at 5\% significance level and 
the results of the spectral analysis 
are given
in Table \ref{timespecfit}. Figure \ref{specvar} shows the variations of
the spectral parameters from one interval to the other in the 
$\nu $-$\nu F_{\nu }$  
representation. The concave shape of the
spectrum of \s5 is apparent.\\
As it can be seen from Table \ref{timespecfit}, 
the emission in the 0.5--10.0 keV band 
becomes most strongly dominated  by the soft synchrotron component during intervals of flaring activity
(i.e. intervals 3 and 5).  The crossing
point
of the two power laws moves to higher energies when the flux increases.
We can also see that the slope of 
the soft power law seems to be anti-correlated with flux, tending
 to get flatter during 
flares. Interestingly, the hard IC
component appears also to vary significantly, but following
 a more complex behavior. 
Up to interval 4, it exhibits a trend similar 
to the one observed
for the soft component, but it keeps steepening during the big flare in
interval 5. However, 
due to the dominance of the synchrotron emission, the slope of the IC 
component in interval 5 is poorly constrained.  \\
We also tested the hypothesis that only one of the two components is varying
with time, whereas the other one is constant. To this purpose, we fitted
the spectra in the various intervals,
fixing both the slope and the normalization of one component to the values
observed in interval 4, during which
both the synchrotron and IC emissions gave comparable contributions (but also
other values were tested). 
The parameters of the second component were left free to vary. 
This model failed to satisfactorily represent the data in all the 
time intervals (reduced $\chi ^{2}\mathrm{s}>1$, prob.$\la 10^{-4}$), 
independently of which component was kept constant. 
  We thus find difficult to
 reconcile the data with the scenario in which only
one component is variable. In particular the data do not appear to 
support the scenario
in which the IC component is stable on short time scales of hours.\\ 
All the above results suggest that the flares are essentially caused by the
high-energy tail of the synchrotron component, which undergoes episodes of
enhanced emission, with simultaneous extension to harder energies. 
 The synchrotron component appears to actually follow
a harder-when-brighter trend, in agreement with that
 observed for the synchrotron-dominated HBL. The apparent softer-when-brighter
 trend observed in Fig. \ref{hr} is thus explained by the relative
 contribution
 of the soft
 synchrotron component, getting higher during flares 
with respect to that of the IC component. This implies  
an overall steepening of the spectrum,
although the actual soft slope becomes
flatter. \\

\begin{table*}
\begin{center}
\small
\tabcolsep1ex
\caption{\label{timespecfit} Results of the time-resolved spectral analysis 
  of the PN+MOS data (see
  Fig. \ref{totlight} for the definition of the time intervals). A
 double power law fit plus Galactic absorption was used. 
  Cols. 1 and 2: photon indices of the two power laws; Col. 3: 
  crossing point; Cols. 4 and 5:
  PN 0.5--10.0 keV fluxes of the steep and
  flat components; Col. 6:  
  relative contribution of the steep-synchrotron component
  to the 0.5--10.0 keV flux. }
\begin{tabular}{c|cccccc}
\noalign{\smallskip} \hline \noalign{\smallskip}
\hline
\multicolumn{1}{c}{Int.} &  \multicolumn{1}{c}{$\Gamma _{1}$}
& \multicolumn{1}{c}{$\Gamma _{2}$} & \multicolumn{1}{c}{$E_{\rm cross}$}
& \multicolumn{1}{c}{$F_{1}$} & \multicolumn{1}{c}{$F_{2}$}
& \multicolumn{1}{c}{Synchr.}\\
\multicolumn{1}{c}{} &  \multicolumn{1}{c}{}
& \multicolumn{1}{c}{} & \multicolumn{1}{c}{(keV)}
& \multicolumn{1}{c}{$10^{-12}$ \ergs} & \multicolumn{1}{c}{$10^{-12}$ \ergs}
& \multicolumn{1}{c}{(\%)}\\
\noalign{\smallskip} \hline \noalign{\smallskip}
\vspace{0.1cm}
1  & 3.20$^{+0.19}_{-0.16}$ & 1.87$^{+0.18}_{-0.25}$ & 2.01 & 
$5.99\pm 1.31$ & $4.26\pm 1.60$ & 58\\
\vspace{0.1cm}
2 & 3.10$^{+0.09}_{-0.11}$ & 1.50$^{+0.13}_{-0.22}$ & 2.46 & 
$5.08\pm 1.10$ & $3.15\pm 1.33$ & 62 \\
\vspace{0.1cm}
3 & 2.95$^{+0.09}_{-0.07}$ & 1.30$^{+0.26}_{-0.27}$ & 3.74 & 
$7.93\pm 1.49$ & $3.01\pm 1.88$ & 72 \\
\vspace{0.1cm}
4 & 3.16$^{+0.42}_{-0.26}$ & 1.68$^{+0.25}_{-0.34}$ & 1.52 & 
$4.28\pm 1.83$ & $4.96\pm 2.04$ & 46 \\
\vspace{0.1cm}
5 & 2.95$^{+0.28}_{-0.14}$ & 1.87$^{+0.37}_{-0.62}$ & 5.34 & 
$13.01\pm 4.24$ & $3.76\pm 1.91$ & 78 \\

\noalign{\smallskip}\hline
\end{tabular}
\medskip

\end{center}
\end{table*}

Previous X-ray observations of some HBL have revealed loop patterns
 in the $\alpha-I$, or
hardness-ratio vs. count rate (HR-CR) plots                
(Takahashi et
 al. 1996; Cui 2004; Ravasio et al. 2004; Brinkmann et al. 2005). 
These loops imply time lags between different energy bands 
(Kirk, Rieger \& Mastichiadis 1998):
soft lags for clockwise loops and hard lags for
counter-clockwise loops. No loops of either kind have been reported so far for
any LBL/IBL. This might be  attributed partly to the insufficient quality of
 the X-ray data,
and partly to the effect of the 
IC component,
which complicates the situation. However, 
B\"ottcher \& Chiang (2002), incorporating the IC component into their
 models, have shown
that loops are to be expected also in the case of LBL/IBL, although with
 complex patterns, which change according to the set of physical parameters
 adopted.  \\
The high quality of
 the XMM data offer the opportunity to check for the presence of phenomena
of spectral hysteresis in the
 case of \s5. 
The cross-correlation analysis for the entire light curve
 (see Sect. \ref{timing}) did not reveal any significant time lag
 between  different energy bands, and thus  no loops should be
 expected in this case.
Since the peak energies of the single flares might be different, loops
 of distinct events may follow different trends and the lags associated with 
them would be averaged in a total cross-correlation function. 
We have thus restricted the search of time lags (and loops)
to
 interesting parts of the light curve, such as 
the two most prominent peaks of the central
burst ($\sim 3.5$ ks each) and the last $\sim 3.5$ ks of the observation, comprising the
small flare superimposed on the bigger
one. In no case significant hysteresis patterns, at more than 3 sigma level,
 could be identified.
However, this analysis does not completely rule out
lags, as only minor flares could be analyzed, whereas
 the largest amplitude events were not fully sampled. A further complication
 is due to the difficulty of isolating single flares, as significant blending
 seems to occur.

\section{Optical data}
\label{optical}

\begin{center}
\begin{figure}
\psfig{figure=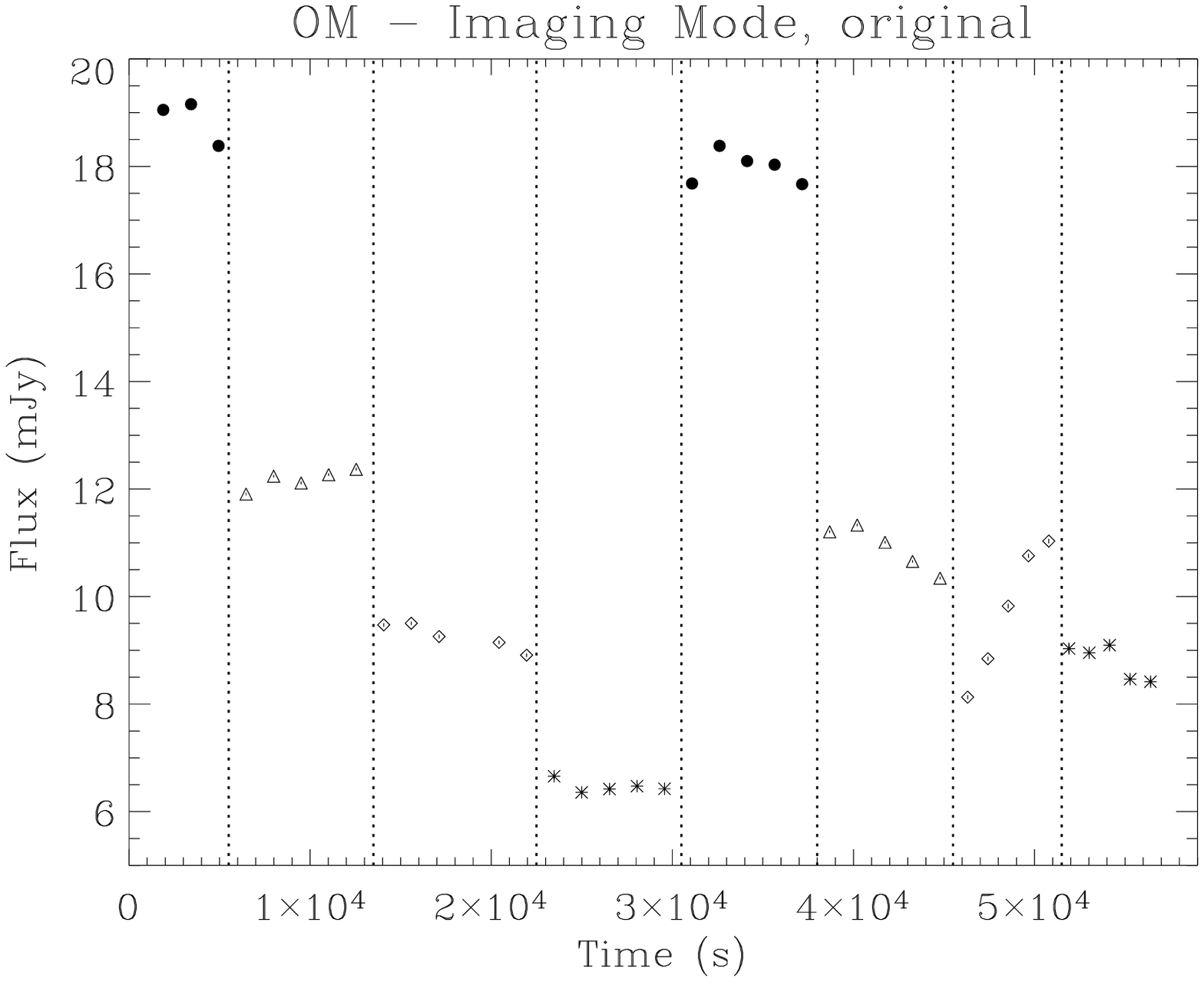,height=3.3truecm,width=7.3truecm,angle=90,%
 bbllx=112pt,bblly=704pt,bburx=453pt,bbury=194pt,clip=}
\psfig{figure=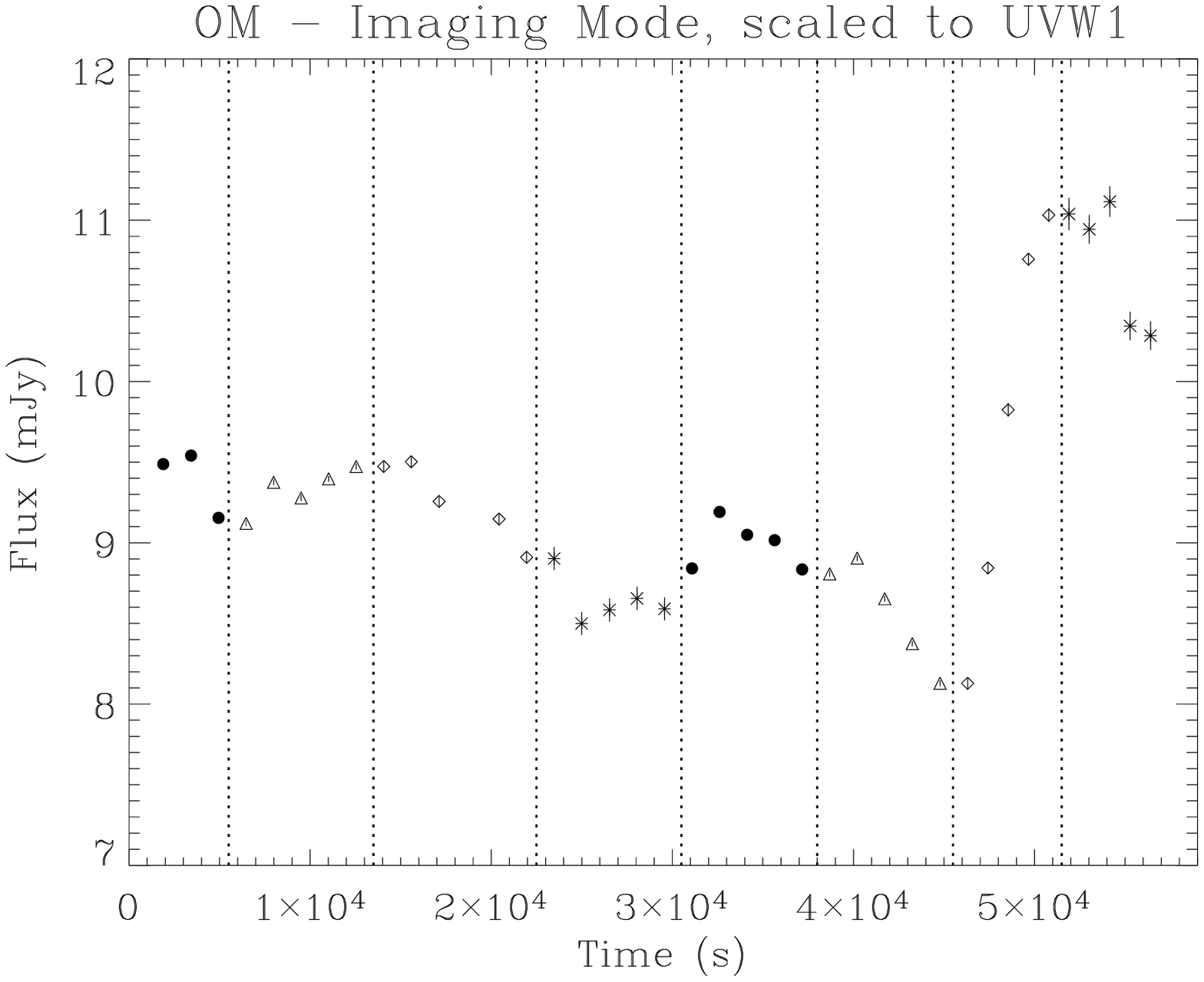,height=3.3truecm,width=7.3truecm,angle=90,%
 bbllx=112pt,bblly=704pt,bburx=453pt,bbury=194pt,clip=}
\psfig{figure=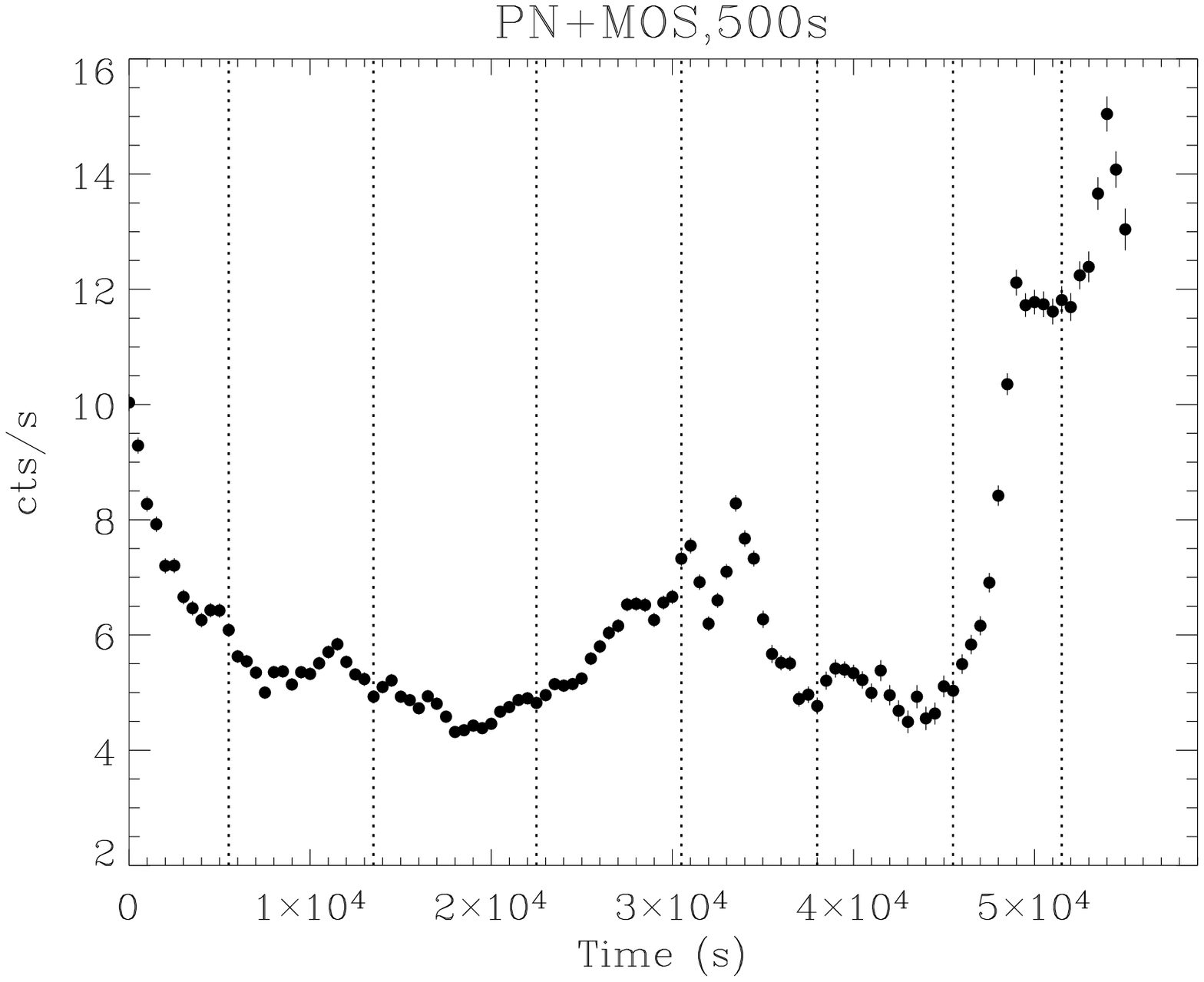,height=3.8truecm,width=7.3truecm,angle=90,%
 bbllx=67pt,bblly=704pt,bburx=453pt,bbury=194pt,clip=}

\caption[]{The original (upper panel) and scaled to the UVW1 band 
(middle panel)
  optical/UV light curves. The different filters are indicated by dots (V),
  triangles (U), diamonds (UVW1) and stars (UVM2). The 0.5--10.0 keV 
light curve
is shown in the bottom panel. The dashed lines separate the
  optical exposures and allow an easier comparison with the X-ray light curve.}
\label{omflux}
\end{figure}
\end{center}

The X-ray data were compared with the optical ones from the
simultaneous observations by the Optical Monitor (OM).
The OM observation consists of a series of 28 exposures in Imaging+Fast mode
lasting $\sim 1200 $ s,
followed by other 10 exposures of $\sim 1000$ s in
Imaging mode only (see Table \ref{obslog}). 
After the first three exposures in the V band, the filter is
changed every five exposures following the sequence V, U, UVW1, UVM2. For each
exposure we obtained an integrated
 flux from the Imaging mode data and a light curve from the Fast mode. 
However, the
 Fast mode light curves from consecutive exposures turned out to be inconsistent
 with each other, with steps of the order of $\sim 1$ mag
 between two exposures
 ($\Delta t\sim 5$ min).  
Therefore, in the
 following we will consider only the data from the Imaging mode.   \\
The OM count rates were converted into fluxes according to the prescriptions 
of the XMM watch-out 
pages\footnote{\tt http://xmm.vilspa.esa.es/sas/new/watchout/}. 
We used the conversion factors for a white dwarf, as recommended by the OM
calibration scientists (Nora Loiseau, priv. comm.). We calculated the extinction for the
various OM
filter wavelengths from the values in the $B_{\rm J}$ band given by
NED (Schlegel et al. 1998) using the algorithms
of Cardelli et al. (1989). The results were used to de-redden the OM 
fluxes. The light curve obtained from this procedure is shown in the top panel
of Fig. \ref{omflux}. \\
In order to enable a more straightforward comparison with the X-ray light
curve, 
we scaled all the optical fluxes to $\lambda =291$ nm,
the central wavelength of the UVW1 filter, assuming a power law
spectrum. For any given filter, the spectral index  
was determined from 
the flux measurements closest in time to those in the UVW1 filter. 
In the case of the V exposures
the fluxes were first scaled to the U band 
and then to the UVW1 band.
The rescaled light curve is shown in
Fig. \ref{omflux} (middle panel) together with the X-ray light curve (lower
panel), for 
comparison. \\
The scaled light curve should be taken with some care, as 
 the optical spectral
index used for the scaling might be affected by significant uncertainty
due to the high degree of spectral variability of the source. 
Furthermore, the variations of the scaled flux might actually not trace
 the true variations in the
UVW1 band. We notice, however, that 
preliminary ground-based optical light curves from a
multi-frequency campaign\footnote{\tt http://www.lsw.uni-heidelberg.de/users/\\
lostorer/0716/0716-nov2003.html}, carried out in the period October 2003--May
2004,
are in overall agreement with our OM scaled light curve. \\
In this paper we restricted our analysis
 to a
general comparison between the OM and
 the X-ray
light curves.
Only the use of higher time resolution, simultaneous
optical/UV data
could serve to firmly constrain the correlation properties 
between the two bands. 
 \\
The most striking feature of the scaled UV
light curve is the big flare at the end of the observation, which traces quite
well that seen in X-rays. The data suggest that the start of the X-ray
flare might occur $\sim 1$ ks before that at UV frequencies. The rise of the 
flux also appears faster in the X-ray band. In general, the amplitude of the
variations seems larger at X-ray energies than at UV frequencies.\\
Apart from the flare at the end, it is hard to identify obvious
correspondences between events in the two light curves. 
Before the final 
flare, the UV 
flux follows a decreasing trend for about 45 ks, with small
amplitude variability
on top of it. There is no clear sign of the middle flare observed in the X-ray
band. The flux decay at the beginning of the X-ray light curve also does not
appear clearly in the UV band. \\
The above results show that, at least for some macroscopic flaring events,
 a correlation exists between the emission in the optical/UV and X-ray bands.

%____________________________________________________________________

\section{Discussion}
\label{discussion}

The analysis of the  XMM observation of \s5 on April 4-5 has revealed 
two spectral
components in the 0.5--10.0 keV band: a soft and steep power law ($\Gamma \sim 3$),
 always dominant
below $\sim 1-2$ keV (depending on the time interval considered), which is
 attributed to
synchrotron emission from the high-energy tail of the electron distribution; a
hard and flat power law ($\Gamma \sim 1.3-1.9$), dominant above $\sim 2-3$ keV, 
associated with the IC
emission of the low-energy tail of the electron distribution. These results
are consistent with previous studies in the X-rays (Cappi et al. 1994; Giommi et al. 1999;
Tagliaferri et al. 2003). The higher sensitivity of XMM
allowed to better constrain the spectral parameters, also in
sub-intervals of the exposure. The X-ray SED of \s5 clearly 
exhibits a concave shape. 
The transition region 
 between the tail of the synchrotron bump and the low-energy
side of the IC peak varies during the observation depending on the spectral
indices of the two components. \\
The source displayed strong variability associated with spectral changes on
time scales of hours. 
The largest variation during the exposure was an
increase in count rate of a factor of $\sim 3$ in $\sim 4$ ks towards
the end of the observation. Such flux variations are not
uncommon for this source and have been reported before (Wagner 1992;
Cappi et al. 1994;
Tagliaferri et al. 2003). The shortest time scale of the flux variations
inferred from the PDS of the 0.5--10.0 keV light curve is $\sim 2.5$ ks,
implying that the emitting region has a size $R\la 2.5\times 10^{-5}
\delta/(1+z)$ pc $\approx 0.7\delta/(1+z)$ light-hours.
Using $\delta \sim 8$ for $z=0.3$ (Ostorero et al. 2006) would result in a
size of the emitting region of  $R\la 4.3$ light-hours;
using  $\delta \sim 24$ (Agudo et al. 2006) would result in a
size of  $R\la 13$ light-hours.\\
The model-independent hardness-ratio analysis 
indicated that during bursts the overall spectrum softens. 
This behavior was already observed with {\it Beppo}SAX 
by Giommi et al. (1999), whereas with ROSAT Cappi et al. (1994) reported a
harder-when-brighter trend. However, in  both cases the observed
tendencies were interpreted in terms of
the synchrotron tail extending to higher energies during high states, while
the IC component remained constant. The apparent inconsistencies of the 
reported behaviors of the hardness-ratios should be ascribed to the 
different energy
ranges of the instruments used. \\
Cappi et al. (1994) were not able to distinguish whether the synchrotron or the IC
component was mostly responsible for the observed variability in the
0.1--2.4 keV band. Giommi et al. (1999) and  Tagliaferri et
al. (2003) argued in favour of a variable synchrotron emission and a stable IC
component, on time scales of hours. This was based on the lack of any
significant variability above $\sim 3-5$ keV, where the IC component is
dominant. With the XMM data, we did find 
from the calculation of FVA
for the light curves in different energy
bands (see Sect. \ref{timing}) that the variability amplitude 
decreases with 
energy. However,  above $\sim 3$ keV, where the
IC component starts to dominate, it is still significantly high 
(FVA$\sim $27\%). 
This change in FVA 
could be 
due to the effect of the smallest contribution of the synchrotron emission
with respect to the IC component.
To test this hypothesis, we assumed, in a very simple scenario,
that the intrinsic variability amplitude of the synchrotron component does not 
depend on
energy within the XMM energy range, and that the observed FVA in 
a given band
simply scales with the relative contribution of the
synchrotron emission to that band 
(i.e. FVA$_{\rm i}\propto F_{\rm i}^{\rm sync}/F_{\rm i}^{\rm tot}$, 
in any given band i) whereas the IC component does not vary.
 Assuming also that in the 0.5--0.75 keV band, 
where the synchrotron emission is
largely dominant ($\sim 89$\%), the observed FVA is approximately equal to the
intrinsic variability amplitude of the synchrotron component, we expect that 
FVA$\sim 13$\% in the 3--10 keV band. 
As one can see, the observed 3--10 keV  FVA 
($\sim 27$\%) is much
higher than the expected one. The reason for that could be either that 
the IC component also gives a contribution  to
the FVA in the hardest band, that 
the FVA 
of the synchrotron component increases with energy,
 or that both effects are at work. 
The
time-resolved
spectral analysis (Sect. \ref{timespectral}) showed that the data are best
modeled with a varying IC component 
on short time
scales. We thus suggest that the IC emission contributes to increase
the variability above $\sim
1$ keV with respect to that expected from simple arguments.
As the
higher-energy electrons cool faster,
the actual
variability amplitude of the synchrotron tail
is indeed expected to increase with energy, and thus 
this effect might also give a contribution to the FVA in the hardest band. \\
The light curves in different energy bands are well
correlated with each other and no significant time lags exceeding $\sim 100$ s 
could be found
between them.
However, there might also be
several effects which might mask the existence of any true delay.
As already mentioned in Sect. \ref{timespectral}, if different flares 
display lags with different signs, the cross-correlation analysis of the
whole light curve, or of parts of it, constituted by a superposition of several
flaring events, might yield an overall zero lag. From the light curve in
Fig. \ref{totlight} the difficulty of isolating single
flares is obvious.
Another possibility for the lack of the detection of lags
 could be that the 
flux variability is set by the
light-crossing time and not by the acceleration or cooling times. In that
case, one would expect to observe flares with rather symmetrical profiles. The
smaller bursts analyzed in Sect. \ref{timespectral} are superimposed on larger
 flares and might as well
be  blended
with other flares,
rendering the characterization of their time profile 
rather problematic. Unfortunately, none of the major bursts in the light
curve, i.e. the ones at the beginning and at the end of the
observation, was fully monitored and the overall profile of these flares could
not be determined either. \\ 
Data from the OM (Sect. \ref{optical}) suggest a soft lag ($\sim 1$ ks) 
between the UV and X-ray bands for the flare at the end of the exposure. 
This soft lag would be in
agreement with the cooling-dominated scenario and with observations of HBL.\\
The results of the spectral and timing analyses of the same
XMM data set by 
Foschini et al. (2006) (see Sect. \ref{intro}) are generally
consistent with ours. On the other hand, 
they concluded 
that the IC component is stable on short time
scales of a few hours, whereas we argued that the IC component varies
also on these time scales. However, for their  time-resolved spectral analysis,
 Foschini et al. used a smaller data sample than ours.
They excluded in particular the last part of the observation, 
covering the most prominent flare of the source, whereas  
our results are based on a detailed investigation of the
variability properties of the  
source over the entire exposure.\\
Foschini et al. compared the EPIC-MOS2  X-ray light curve
with that in the V band from 
the INTEGRAL Optical Monitor Camera, which observed \s5 quasi-simultaneously 
to XMM. 
No correlation could be established between the two light curves,
 because of different
samplings and time gaps. By means
of the
scaled XMM-OM optical/UV light curve,
we revealed a generally good 
correlation 
with the X-ray band.

%_____________________________________________________________________

\section{Conclusions}
\label{conclusions}

In this paper we have reported the results from the data analysis of a $\sim
$59 ks XMM observation of the BL Lac object \s5. The shape of the spectrum of
the source within 0.5--10 keV was well constrained. A sum of two power laws, a
steep one at soft energies plus a flat one at hard energies, 
represented the data satisfactorily. The soft power law was related to
synchrotron emission of the high-energy electrons, whereas the hard power law
was interpreted as IC emission from the low-energy electrons. No emission or
absorption feature was detected.\\
Both the synchrotron and the IC components
appeared to vary on time scales of hours. The synchrotron emission shows the 
largest
variability amplitude.
It was found
to flatten and extend towards higher energies during high states. 
For the IC component 
 no clear
correlations of the spectral variations with the flux
level could be established. The data are not compatible with 
the hypothesis that the IC component is constant on short time scales, as
claimed previously.   \\
The 0.5--10.0 keV light curve displayed strong and fast variability, with the
largest variation occurring towards the end of the observation, distinguished
by an increase in count rate of a factor $\sim 3$ in $\sim 4$ ks.  
The inferred size of the emitting region is $R\la 0.7\delta /(1+z)$ 
light-hours. The
variability in different energy bands is well correlated, excluding delays
$\ga 100$ s. The absence of any significant lag between synchrotron 
radiation emitted from the high-energy end of the electron distribution
and the IC emission resulting from scattering off low-energy electrons
suggests that there are no significant lags across the energy spectrum
of the underlying particle distribution. \\

\begin{acknowledgements}
This work is based on observations with XMM-Newton, an ESA science mission
with instruments and contributions directly funded by ESA Member States and
the USA (NASA). This research has made use of the NASA/IPAC Extragalactic 
Database (NED), which is operated by the Jet Propulsion 
Laboratory, California Institute of Technology, 
under contract with the National Aeronautics and Space Administration. 
We acknowledge support by BMBF, through its agency DLR for the 
	project 50OR0303 (S. Wagner). We 
acknowledge EC funding under contract HPRN-CT-2002-00321 (ENIGMA). We thank
M. Freyberg for useful discussions about the XMM data analysis and
I. Papadakis for writing the FORTRAN routine for the cross-correlation
analysis and for help with the timing analysis. 
\end{acknowledgements}

\end{document}